\documentclass[%
 preprint,
 amsmath,amssymb,
 aps,
 pra,
]{revtex4-1}

\usepackage{graphicx}
\usepackage{dcolumn}
\usepackage{bm}
\usepackage{color}

\begin{document}

\preprint{APS/123-QED}

\title{Measuring the continuous variable quantum entanglement with a parametric amplifier assisted homodyne detection}
\author{Jiamin Li$^{1}$}
\author{Yuhong Liu$^{1}$}
\author{Nan Huo$^{1}$}
\author{Liang Cui$^{1}$}
\author{Sheng Feng$^{2}$}
\author{Xiaoying Li$^{1}$}
 \email{xiaoyingli@tju.edu.cn}
\author{Z. Y. Ou$^{1, 3}$}
\affiliation{%
$^{1}$College of Precision Instrument and Opto-Electronics Engineering, Key Laboratory of
Opto-Electronics Information Technology, Ministry of Education, Tianjin University,
Tianjin 300072, P. R. China\\
$^{2}$Beijing Institute of Radio Measurement, Beijing, P. R. China\\
$^{3}$Department of Physics, Indiana University-Purdue University Indianapolis, Indianapolis, IN 46202, USA
}%

\date{\today}

\begin{abstract}
Traditional method for measuring continuous-variable quantum entanglement relies on balanced homodyne detections, which are sensitive to vacuum quantum noise coupled in through losses resulted from many factors such as detector's quantum efficiency and mode mismatching between detected field and local oscillator. 
In this paper, we propose and analyze a new measurement method, which is realized by assisting the balanced homodyne detections with a high gain phase sensitive parametric amplifier. 
The employment of the high gain parametric amplifier helps to tackle the vacuum quantum noise originated from detection losses.
Moreover, because the high gain parametric amplifier can couple two fields of different types in a phase sensitive manner, the proposed scheme can be used to reveal quantum entanglement between two fields of different types by using only one balanced homodyne detection. 
Furthermore, detailed analysis shows that in the multi-mode case, the proposed scheme is also advantageous over the traditional method. Such a new measurement method should find wide applications in quantum information and quantum metrology involving measurement of continuous variables.
\end{abstract}
\maketitle


\section{Introduction}
Homodyne detection, which measures the quadrature amplitudes of light, plays a central role in continuous variable (CV) quantum information processing (QIP), from characterizing the CV entanglement to fulfilling various kinds of QIP tasks \cite{bra00,zh00,li02,snb13,bennett93,bou97,vaid94,bra98,furu98,dowl00,gio06,ligoepr}. 
In particular, the measurement of CV entanglement between two parties usually relies on the joint measurement of two sets of balanced homodyne detectors (BHD), in which a 50/50 beam splitter is used to superpose the optical input and the local oscillator (LO). 
However, the measurement performed by BHD is prone to losses such as propagation loss, less-than-unit quantum efficiency of detectors and imperfect mode matching efficiency between the signal input and LO. 
As a result, the observed quantum effect is smaller than what is anticipated, and the quantum advantage of entanglement is often hampered by detection losses. 
Moreover, BHD with high efficiency for some fields such as 2 $\mu$m optical wave and atomic wave are not available yet, so the investigation and application of entangled fields with wavelengths out of the mature detection technologies are obstructed.

Recent investigation shows that a high gain phase sensitive amplifier (PSA) can be viewed as homodyne detection, in which the strong pump of PSA serves as the LO. In Ref. \cite{Shaked-NC}, the squeezing generated from four-wave-mixing (FWM) in fiber was successfully measured by scanning the pump phase of PSA, and the results illustrated that the noise reduction measured by PSA have the advantageous of detection loss tolerance. Indeed, the idea of using high gain PSA to measure quantum noise reduction was first adopted by Flurin {\it et al.} to overcome the huge classical electronic noise in measuring the entanglement of microwaves \cite{Flurin-PRL}, in which the PSA formed by Josephson mixer functions as a dis-entangler. The results in Refs. \cite{Shaked-NC} and \cite{Flurin-PRL} show that the inseparability of entanglement between two fields can be characterized by directly measuring the intensity at one output of PSA. 
This kind of method has two advantages. One is the tolerance to detection loss, the other is that only the measurement of one field is required. However, comparing with the traditional method of measuring entanglement with two BHDs, the method of using PSA followed with a power detector for measuring noise reduction has the following deficiencies.  First, the measured noise fluctuations of quantum states in Refs. \cite{Shaked-NC} and \cite{Flurin-PRL} are determined by the power of PSA output, so the gain of PSA should be very high. 
The high pump power required for achieving the high gain hinder the practicality to a certain extent. Second, the intensity directly measured at one output of PSA is the sum effect of noise for both the quadrature amplitudes $\hat X_i$ and $\hat Y_i$ ($i=$ 1, 2) of two entangled fields, but the noise fluctuation of the quadrature amplitudes of two entangled fields, $\hat X_1 \mp \hat X_2$ and $\hat Y_1 \pm \hat Y_2$, can not be respectively measured. Therefore, this method, which can characterize the inseparability of entanglement, is not suitable for fulfilling many QIP tasks, such as quantum dense coding \cite{bra00,zh00,li02,snb13}, teleportation \cite{bennett93,bou97,vaid94,bra98,furu98}, and quantum enhanced precision measurement \cite{dowl00,gio06,ligoepr}.

In this paper, we propose and analyze a new method realized by assisting the BHDs with a PSA. The new method inherits the advantages of both BHDs and high gain PSA in measuring CV entanglement. In addition to the tolerance to detection loss and measuring entanglement between two fields by detecting only one field, the reduced noise fluctuations due to the quantum correlation between two entangled fields, $\langle \Delta^2 (\hat X_1 \mp\hat X_2) \rangle$ and  $\langle \Delta^2 (\hat Y_1 \pm \hat Y_2) \rangle$, are respectively and simultaneously measurable. Moreover, by studying the dependence of measurement upon the key parameters of the new scheme,  we find these advantages can be achieved by using a PSA with moderate gain, which makes the new method more practical.

The rest of the paper is organized as follows. In Sec. II, we briefly review the traditional method of measuring entanglement by using the joint measurement of two BHDs. In Sec. III, we introduce the new type homodyne detection realized by a high gain PSA followed with a detector for directly measuring the intensity at one output of PSA. 
In Sec. IV, we investigate the new measurement method realized by using PSA to assist the joint measurement of two BHDs and individual homodyne detection of one BHD, respectively. 
The simulation results obtained by varying the gain of PSA and the loss of BHD clearly demonstrate that the performance of new method surpasses those schemes in Secs. II and III, and the traditional method in Sec. II is just a specific case of the new method for PSA with pump power turning off. 
In Sec. V, we extend the measurement of entangled states from the single mode model to multi mode model. The detailed analysis shows that the new method is also advantageous over the traditional method in multi-mode case. Finally, we conclude in Sec. V.

\section{Traditional method for measuring entanglement}


Einstein-Podolsky-Rosen (EPR) type entangled state was first put forward by Einstein, Podolsky, and Rosen in Ref. \cite{epr}. It says that there exists such a state of two particle systems that exhibits perfect correlations not only between their positions but also between their momenta.
This is so because the difference of their position operators $\hat x_1 - \hat x_2$ and the sum of their momenta operators $\hat p_1 + \hat p_2$ commute: $[\hat x_1 - \hat x_2, \hat p_1 + \hat p_2]=0$.
The experimental realization of the EPR state and the demonstration of EPR paradox were first done in an optical system of non-degenerate parametric amplifier (PA) \cite{reid89,ou92} in which the two particles are the virtual harmonic oscillators representing two spatially separated modes of optical beams with $\hat x_{1,2} \propto \hat a_{1,2}^{\dag} + \hat a_{1,2} \equiv \hat X_{1,2}$ and $\hat p_{1,2} \propto -i(\hat a_{1,2}^{\dag} - \hat a_{1,2})\equiv \hat Y_{1,2}$, where $\hat a_{i}^{\dag}$ and $\hat a_{i}$ ($i=1,2$) are the creation and annihilation operators of the two optical fields, $\hat X_i$ and $\hat Y_i$, are the quadrature-phase amplitude $\hat X(\phi) = \hat a e^{-i\phi} + \hat a^\dag e^{i \phi}$ with $\phi=0$ and $\pi/2$, respectively.
For the conjugate observables $\hat X_i$ and $\hat Y_i$ ($i=1,2$), although the commutation relation  $[\hat X_i, \hat Y_j]=\delta_{i,j}$ holds, quantum mechanics permits the commutation relation: $[\hat X_1 -\hat X_2, \hat Y_1+\hat Y_2]=0$, which means the quantum fluctuations for the difference and sum of quadrature amplitudes, $\hat X_1-\hat X_2$ and $\hat Y_1+\hat Y_2$, can simultaneously approach to zero.
For the sake of convenience, in this paper, we will take the most common source of EPR state realized by a phase insensitive PA as an example.

\subsection{EPR states generated from parametric amplifiers}

\begin{figure}[htb]
\centering
 \includegraphics[width=4.5cm]{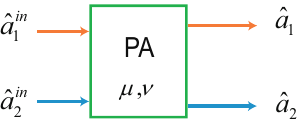}
\caption{Entangled state generated from a parametric amplifier (PA).}
\label{fig-EPR-source}
\end{figure}

Figure \ref{fig-EPR-source} shows the schematic of the entanglement source.
The two input and output fields of PA are labeled as $\hat a_1^{in}, \hat a_2^{in}$ and $\hat a_1, \hat a_2$, respectively, and the input fields are both in vacuum.
Ideally, the input-output relations of the PA are given by
\begin{equation}\label{Gg}
\begin{split}
\hat a_1 &= \mu \hat a_1^{in} + \nu \hat a_2^{{in}\dag},\\
\hat a_2 &= \mu \hat a_2^{in} + \nu \hat a_1^{{in}\dag},
\end{split}
\end{equation}
where $\mu$, $\nu$ with $\mu^2-\nu^2=1$ are the amplitude gains of the PA.
The noise variances for the difference and sum of quadrature amplitudes at the two outputs, $\hat X_- = \hat X_1 -\hat X_2$ and $\hat Y_+ = \hat Y_1 + \hat Y_2$, are expressed as
\begin{equation}\label{noise_EPR}
\begin{split}
	\langle \Delta^2 \hat X_- \rangle&=\langle \Delta^2 (\hat X_1 -\hat X_2) \rangle=2(\mu-\nu)^2,\\
	\langle \Delta^2 \hat Y_+ \rangle&=\langle \Delta^2 (\hat Y_1 +\hat Y_2) \rangle =2(\mu-\nu)^2.
\end{split}
\end{equation}
Defining the vacuum noise of one optical field as 1, the corresponding shot noise limits (SNL)
\begin{equation}\label{SNL_EPR}
 \langle \Delta^2  \hat X_-  \rangle_{SNL}=
 \langle \Delta^2 \hat Y_+  \rangle_{SNL} =2
 \end{equation}
can be obtained by replacing the two fields $\hat a_1, \hat a_2$ with vacuum. Since the variances of $\hat X_-$ and $\hat Y_+$ in Eq. (\ref{noise_EPR}) are lower than those in Eq. (\ref{SNL_EPR}), after normalizing Eq. (\ref{noise_EPR}) with the corresponding SNLs in Eq. (\ref{SNL_EPR}), we arrive at the noise reduction of the entangled source
\begin{equation}\label{EPR-Nor}
\begin{split}
   \langle \Delta^2 \hat X_- \rangle_{s}=\frac{\langle \Delta^2 \hat X_- \rangle}{\langle \Delta^2  \hat X_-  \rangle_{SNL}} =
 (\mu-\nu)^2<1,\\
  \langle \Delta^2 \hat Y_+ \rangle_{s}=\frac{\langle \Delta^2 \hat Y_+ \rangle}{\langle \Delta^2  \hat Y_+  \rangle_{SNL}} =
 (\mu-\nu)^2<1, 
\end{split}
\end{equation}
Accordingly, the inseparability criterion \cite{duan00}
 \begin{equation}
  I_s \equiv  \langle \Delta^2 \hat X_- \rangle_{s} + \langle \Delta^2 \hat Y_+ \rangle_{s} =  2(\mu-\nu)^2 < I_s^{SNL}=2
 \label{I_s}
 \end{equation}
for $\nu \neq 0$, where $I_s^{SNL}=2$ denotes the SNL of the inseparability coefficient. The smaller $I_s$ is, the higher the entangled degree between the two fields is. For the case of $\nu \rightarrow \infty$, we have $I_s \rightarrow 0$, which means the two fields $\hat a_1$ and $\hat a_2$ are perfectly correlated and completely entangled if the gain of PA is very high.

\subsection{Traditional method for the entanglement measurement}

\begin{figure}[htb]
	\centering
	\includegraphics[width=7cm]{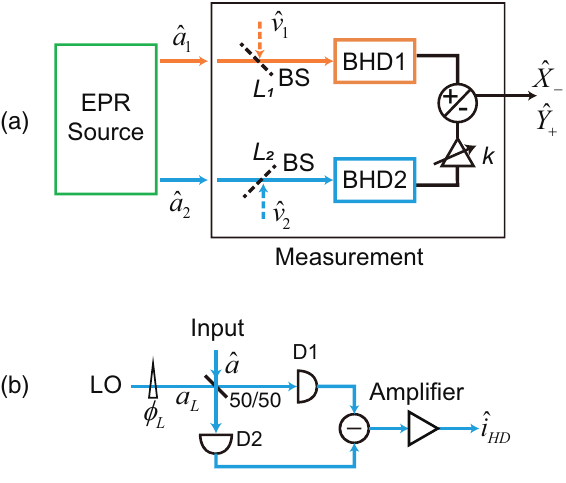}
	\caption{(a) Traditional method for Einstein-Podolsky-Rosen (EPR) type entanglement measurement. 
	(b) The scheme of a balanced homodyne detection (BHD) system.
	D1-D2, detectors; LO, local oscillator.
	The detection loss of field $\hat a_1/\hat a_2$ is modeled as beam splitter (BS) of reflectivity $L_1/L_2$. $\hat v_1/\hat v_2$ represents the vacuum field coupled into the detected fields.}
	\label{fig-tri-method}
\end{figure}

Traditionally, the EPR entangled state is characterized by the joint measurement of two BHDs. As shown in Fig. \ref{fig-tri-method}(a), the quadrature-phase amplitudes of the two entangled fields are respectively measured by BHD1 and BHD2. The configuration of each BHD is shown in Fig. \ref{fig-tri-method}(b).
Each BHD consists of a 50/50 beam splitter (BS), two detectors (D1 and D2), and a subtracter followed with an electrical amplifier. The BS is used to superpose the local oscillator (LO) with the weak signal input field $\hat a$. 
The LO is a strong coherent light $|\alpha_{L} \rangle = |\alpha_{L}|e^{i\phi_L}$, where the amplitude $|\alpha_{L}|$ is much higher than that of the signal input ($|\alpha_L| \gg 1$), and $\phi_L$ represents the phase of LO. The two outputs of BS are respectively measured by D1 and D2. The difference between the photo-currents of D1 and D2 is $\hat i \propto |\alpha_{L}|\hat X(\phi)$, 
where $\phi$ can be varied by changing the phase of LO.
When the phase $\phi$ takes the value of 0 and $\pi/2$, the photo-current operator $\hat i$ corresponds to measurement of quadrature amplitudes $\hat X =\hat X(0)$ and $\hat Y =\hat X(\pi/2)$, respectively. 
The AC component of $\hat i$ is amplified to get rid of the influence dark current and the variance of the photo-current out of one BHD is
\begin{equation}
\label{HD-noise}
	\langle \Delta^2 \hat i_{BHD}\rangle = q^2 |\alpha_{L}|^2 \langle \Delta^2 \hat X(\phi)\rangle,
\end{equation}
where $q$ is the gain of electronic amplifier.

The joint measurement of two fields (see Fig. \ref{fig-tri-method}(a)) is performed by using an electronic combiner. The subtraction and addition operation of the photo-currents of two BHDs give
\begin{equation}
\label{HD-joint}
\begin{split}
    \hat i_{-} &= \hat i_1 - k\hat i_2\propto \hat X_1 -k\hat X_2,\\
    \hat i_{+} &= \hat i_1 + k\hat i_2 \propto \hat Y_1 +k\hat Y_2,
\end{split}
\end{equation}
when the phase $\phi$ of each BHD is locked at $0$ and $\pi/2$, respectively, where $k$ is the adjustable gain for optimizing the measurement. The value of $k$ is usually related to the symmetric property of entanglement and the detection efficiency of each BHD \cite{ou92}. 
The variances of $\hat i_{-}$ and $\hat i_{+}$ for $k=1$ are
\begin{equation}\label{current-jm}
\begin{split}
    \langle \Delta^2 \hat i_-\rangle = q^2 |\alpha_L|^2 \langle \Delta^2 \hat X_-\rangle,\\
    \langle \Delta^2 \hat i_+\rangle = q^2 |\alpha_L|^2 \langle \Delta^2 \hat Y_+\rangle.
\end{split}
\end{equation}
Normalizing Eq. (\ref{current-jm}) with the corresponding SNLs
\begin{equation}
\label{current-SNL}
\begin{split}
    \langle \Delta^2 \hat i_-\rangle_{SNL} = q^2 |\alpha_L|^2 \langle \Delta^2 \hat X_-\rangle_{SNL},\\
    \langle \Delta^2 \hat i_+\rangle_{SNL} = q^2 |\alpha_L|^2 \langle \Delta^2 \hat Y_+\rangle_{SNL},
\end{split}
\end{equation}
we arrive at
\begin{equation}
\label{current-jm-nor}
\begin{split}
    \frac{\langle \Delta^2 \hat i_-\rangle}{\langle \Delta^2 \hat i_-\rangle_{SNL}} = \frac{\langle \Delta^2 \hat X_-\rangle}{\langle \Delta^2 \hat X_-\rangle_{SNL}}=\langle \Delta^2 \hat X_-\rangle_s=(\mu-\nu)^2,\\
    \frac{\langle \Delta^2 \hat i_+\rangle}{\langle \Delta^2 \hat i_+\rangle_{SNL}} = \frac{\langle \Delta^2 \hat Y_+\rangle}{\langle \Delta^2 \hat Y_+\rangle_{SNL}}=\langle \Delta^2 \hat Y_+\rangle_s=(\mu-\nu)^2.
\end{split}
\end{equation}
Equations (\ref{current-jm})-(\ref{current-jm-nor}) clearly illustrate that the joint measurement of two BHDs gives the noise reduction of the source $\langle \Delta^2 \hat X_- \rangle_{s}$ and $\langle \Delta^2 \hat Y_+ \rangle_{s}$ expressed in Eq. (\ref{EPR-Nor}), and it is straight forward to verify the inseparability of the EPR source $I_s =\langle \Delta^2 \hat X_- \rangle_{s}+\langle \Delta^2 \hat Y_+ \rangle_{s} = 2(\mu-\nu)^2<2$ (see Eq. (\ref{I_s})).

\subsection{Challenges of the traditional measurement method}

From Eqs. (\ref{HD-noise}) and (\ref{current-jm})-(\ref{current-jm-nor}), one sees that
the role of strong LO in BHD is to increase the photo-currents of photo-detectors and to effectively amplify a weak input signal field to a level that is otherwise buried in the classical noise such as detector's dark current, thermal electronic noise of current amplifiers and ambient background light. However, the BHD cannot do much against vacuum noise coupled in through loss inevitably existed in a real system, including less-than-unit quantum efficiency, non-perfect propagation and mode mismatching.

We can model the detection loss of BHD1 and BHD2 as BSs with reflectivity $L_1$ and $L_2$ (see Fig. \ref{fig-tri-method}(a)), respectively.
The operators of the fields propagating though the BSs are given by: $\hat a'_{1,2} = \sqrt{1-L_{1,2}} \hat a_{1,2}  + \sqrt{L_{1,2}} \hat v_{1,2} $, where $\hat v_{1,2}$ are the vacuum fields coupled into the detected fields through the BS's unused ports.
For the sake of simplicity, we assume the detection losses of two BHDs are equal, i.e., $L_1=L_2=L_D$.
When the operators $\hat a_{1, 2}$ in Eq. (\ref{Gg}) are replaced with $\hat a'_{1,2}$, after some algebra, the noise variance for difference and sum of quadrature amplitude of EPR state is then expressed as
\begin{equation}
	\langle \Delta^2 \hat X'_- \rangle=
	\langle \Delta^2 \hat Y'_+ \rangle =2(1-L_D)(\mu-\nu)^2+2L_D.
\end{equation}
Since the SNL is still the same as Eq. (\ref{SNL_EPR}), it is straightforward to find the expression of the measured inseparability:
\begin{equation}\label{I-loss}
I'=2(1-L_D)(\mu-\nu)^2+2L_D. 
\end{equation}
Under the perfect detection efficiency $L_D=0$, we have $I'=2(\mu-\nu)^ 2$, which is same as Eq. (\ref{I_s}). Otherwise, the measured inseparability of the EPR source is $I'>2(\mu-\nu)^2$. 
For example, for the case of $\mu\rightarrow \infty$, we have $I' \to 2L_D$, which means that the losses are the factors limiting how small the measured inseparability $I'$ can reach.
So the entanglement is very fragile and vulnerable to losses, and the measured degree of entanglement will be degraded due to the existence of detection loss.

In addition to the loss induced degradation on the measured entangled degree, there are other problems. 
First, the requirement of the availability of two BHDs with higher detection efficiency for measuring two different fields restricts the measurement technique to a certain extend. It is challenge to measure entanglement between two fields with different type. For example, the efficient detectors for atom and 2 $\mu$m optical field are not available yet. 
Second, it is difficult to measure the broadband entanglement generated by single pass PA, whose noise reduction is in principle the same for all the frequency band \cite{Slusher87,Kumar90,Wenger04,Eto08,Guo16}. Since the response bandwidth of BHD is limited by the fixed value of product between the gain of electronic amplifier and gain bandwidth (see Fig. \ref{fig-tri-method}(b)), the high the gain of amplifier is, the narrow the gain bandwidth is.
Third, for multi-mode entangled state generated by pulse pumped PA, noise contributed by the mode mismatching between LOs and detected fields might be much larger than the vacuum noise due to the thermal nature of individual field of entanglement, which makes measured degree of pulsed entanglement smaller than what is anticipated \cite{Wasilewski06,guo15}.

\section{Homodyne detection realized by the combination of high gain PSA and a power detector}
The function of homodyne detection can also be realized by a high gain PSA, whose strong pump serves as local oscillator~\cite{Flurin-PRL,Shaked-NC}. Generally speaking, the PSA can be classified into two types. One is the degenerate PSA, in which the modes of the two inputs/outputs are the same; the other is non-degenerate PSA, in which the modes of two inputs/outputs are different. In this section, we will analyze the performance of such a new type homodyne detector in measuring noise reduction of optical fields when the PSA is degenerate and non-degenerate, respectively.


\subsection{Degenerate phase sensitive amplifier for measuring the noise fluctuation of an optical field}

\begin{figure}[htb]
	\centering
	\includegraphics[width=6.5cm]{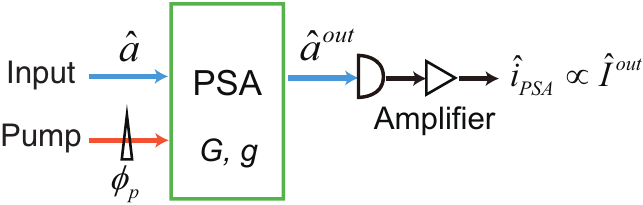}
	\caption{The scheme of measuring the noise fluctuation of input field by using a degenerate high gain phase sensitive amplifier (PSA) followed with a power detector.}
	\label{fig-new-type}
\end{figure}


Figure \ref{fig-new-type} shows the schematic diagram of measuring the noise fluctuation of input field $\hat a$ with a degenerate PSA. The output of PSA is directly measured by a power detector. The strong pump and weak input field are coherently combined via the nonlinear interaction in PSA.
The relationship between the input and output fields of PSA is
\begin{equation}
\label{PSA_squeezed}
	\hat a^{out} =G \hat a + g e^{i\varphi}\hat a^\dag,
\end{equation}
where $\hat a$ and $\hat a^{out}$ respectively represent the input and output fields, $G,g$ with $G^2-g^2=1$ is the gain amplitude of PSA mainly determined by the pump power, $\varphi$ is the difference of phases between the pump and input fields.
The intensity operator of the output field $\hat a^{out}$ is
\begin{equation}
\begin{split}
\hat I^{out} &= \hat  a^{out\dag} \hat  a^{out}\\
&= G^2  \hat  a^{\dag} \hat  a + g^2  \hat a \hat  a^{\dag}+ Gg e^{i \varphi} \hat a^{\dag} \hat  a^{\dag} +Gg e^{- i \varphi} \hat a \hat  a.
\end{split}
\end{equation}

The detector measures the average of the output intensity. If the input is in squeezed vacuum state \cite{Shaked-NC}, the average intensity at the output of PSA is written as
\begin{equation}\label{Int_all-terms}
\begin{split}
    \langle \hat I^{out} \rangle
    &= \frac{(G+g)^2}{4} \langle \Delta^2 \hat X^2(\varphi)\rangle\\
    &\quad+\frac{(G-g)^2}{4} \langle \Delta^2 \hat X^2(\varphi+\frac{\pi}{2}) \rangle  -\frac{1}{2}.
\end{split}
\end{equation}
Equation (\ref{Int_all-terms}) shows the intensity is related to the noise of quadrature amplitude of input field. When the gain of PSA is very high and the first term dominate, we then have the approximation
\begin{equation}\label{noise_degenate}
\langle \hat I^{out}\rangle = g^2\langle \Delta^2\hat X(\varphi) \rangle.
\end{equation}
Equation (\ref{noise_degenate}) clearly illustrate that the noise of the quadrature amplitude of input field $\langle \Delta^2\hat X(\varphi) \rangle$ is amplified by the gain of PSA, and the phase difference $\varphi$ can be changed by changing the phase of pump. For example, for the $\chi^{(2)}$ crystal based PSA, we have $\varphi = \phi_{p} - 2\phi_{in} $, where $\phi_{p}$ and $\phi_{in} $ are the phases of pump and input fields, respectively. The noises of two conjugated amplitudes of input, $\langle \Delta^2\hat X\rangle $ and $\langle \Delta^2\hat Y\rangle $, can be obtained by setting the phase $\varphi$ at $0$ and $\pi$, at which the PSA is operated at the amplification and de-amplification condition, respectively.
In order to characterize whether the noise is lower than vacuum, we need compare the noise of input field with the corresponding SNL,
\begin{equation}\label{SNL-PSA}
	I^{out}_{SNL} = g^2,
\end{equation}
which is obtained by replacing the input field of PSA with vacuum. If the normalized intensity $\langle \hat I^{out} \rangle/I^{out}_{SNL}$ is lower than one, the noise of input field $\langle \Delta^2\hat X(\varphi) \rangle$ is squeezed \cite{Shaked-NC}.

Although the noise of input field is amplified by the gain factor $g^2$, in practice, the gain of PSA can not be infinitely large. To observe the output intensity of PSA, it is necessary to pass the photo-current of detector through a electrical amplifier, whose output is expressed as
\begin{equation}\label{noise_degenate-amp}
\langle \hat i_{PSA} \rangle =q^2\langle \hat I^{out}\rangle = q^2 g^2\langle \Delta^2\hat X(\varphi) \rangle,
\end{equation}
where $q$ is the gain of electrical amplifier. The comparison between Eqs. (\ref{noise_degenate-amp}) and (\ref{HD-noise}) shows that the parametric gain provided by strong pump of PSA functions as the LO of new type homodyne detection \cite{Shaked-NC}.

It is worth emphasizing the approximation of Eq. (\ref{noise_degenate-amp}) only holds when the gain PSA is very high. This is because that in Eq. (\ref{Int_all-terms}), $g$ should be large enough
to ensure the first term dominates and the other terms are negligibly small. It's easy to make the second term small enough but it is not trivial to ensure the third term (originated from the commutation relation $[\hat a^{out\dag},~\hat a^{out}]=1$) is negligible.
For example, for the squeezed state input with $\langle \Delta^2 \hat X(\varphi)\rangle =0.25$, the second term is more than a hundred times smaller than the first term for $g>3$. However, to make the third term,  50 times smaller than the first term, $g>10$ is required.

\subsection{Non-degenerate phase sensitive amplifier for measuring the inseparability of two optical fields}

\begin{figure}[htbp]	
	\centering
	\includegraphics[width=6.5cm]{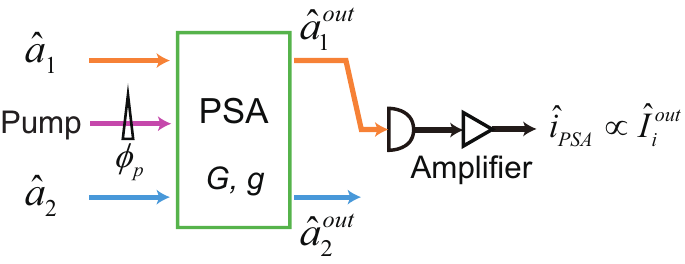}
	\caption{Measuring the inseparability of two fields by using a non-degenerate high gain phase sensitive amplifier (PSA) followed with a power detector.}
	\label{fig-EPR-PA}
\end{figure}



Figure \ref{fig-EPR-PA} shows the scheme for measuring the noise reduction of two fields by using a non-degenerate PSA, followed with a power detector. The modes of the two input fields of PSA, $\hat a_1, \hat a_2$, are non-degenerate, and the relation between the inputs and outputs of the PSA is:
\begin{equation}\label{Gg-opa}
\begin{split}
\hat a_1^{out} &= G \hat a_1 + e^{i \varphi} g \hat a_2^\dag ,\\
\hat a_2^{out} &= G \hat a_2 + e^{i \varphi} g \hat a_1^\dag,
\end{split}
\end{equation}
where $\hat a_1^{out}$ and $\hat a_2^{out} $ are the two output fields, $\varphi$ is the difference between the pump phase and the sum of two input fields' phase. 
Comparing Eq. (\ref{Gg-opa}) with Eqs. (\ref{PSA_squeezed}) and (\ref{HD-joint}), one sees that the gain amplitude $G,g$ of the non-degenerate PSA not only amplifies the two inputs, but also coherently combine $\hat a_1(\hat a_2)$ and $\hat a_2^\dag(\hat a_1^\dag)$ which is similar to the joint measurement of combining $\hat X_1$ and $\hat X_2 $ with ${g}/{G}$ equivalent to $k$ in Eq. (\ref{HD-joint}). So it is reasonable to expect one output of non-degenerate PSA can be used to characterize the entanglement between fields $\hat a_1$ and $\hat a_2$.

At one output field of PA, says $\hat a_1^{out}$, the intensity operator is
\begin{equation}
\begin{split}
&\hat I_1^{out} = \hat  a_1^{out\dag} \hat  a_1^{out}\\
& = G^2  \hat  a_1^{\dag} \hat  a_1 + g^2  \hat  a_2 \hat  a_2^{\dag} + Gg e^{i \varphi} \hat a_2 \hat  a_1 +Gg e^{- i \varphi} \hat  a_1^{\dag}\hat a_2^{\dag}.\\
\end{split}
\end{equation}
The photo-current intensity of power detector measures the average intensity of the field, $\langle \Delta^2 \hat I_1^{out}\rangle$, can be written as
\begin{widetext}
\begin{equation}	\label{intensity_PA}
\begin{split}
\langle \hat i_{PSA} \rangle &= q^2 \langle \hat I_1^{out} \rangle\\
&= q^2\left\{\frac{(G+g)^2}{16} \left[ \langle \Delta^2 (\hat X_1(\varphi)+\hat X_2(\varphi))\rangle\right.
+\langle \Delta^2 ( \hat X_1(\varphi+\frac{\pi}{2}) -\hat X_2(\varphi+\frac{\pi}{2}) ) \rangle \right] \\
& \quad +\frac{(G-g)^2}{16}\left[ \langle \Delta^2 ( \hat X_1(\varphi)-\hat X_2(\varphi) ) \rangle
+\left.\langle \Delta^2 ( \hat X_1(\varphi+\frac{\pi}{2})+\hat X_2(\varphi+\frac{\pi}{2}) )\rangle\right] -\frac{1}{2} \right\}
\end{split}
\end{equation}
\end{widetext}
if the average intensity of each input is the same as vacuum, where $q$ is the gain of electrical amplifier.
Similar to Eq. (\ref{Int_all-terms}), when the gain of PSA is very high, the contribution of the first term in  brace dominates and Eq. (\ref{intensity_PA}) can be approximated as
\begin{equation}\label{Int_g_inf}
\begin{split}
&\langle \hat i_{PSA}\rangle= q^2 \langle \hat I_1^{out} \rangle\\
& \quad= q^2 \frac{(G+g)^2}{16} \left[ \langle \Delta^2 \hat X_+(\varphi)\rangle +\langle \Delta^2 \hat X_-(\varphi+\frac{\pi}{2})\rangle \right].
\end{split}
\end{equation}

To better understand Eq. (\ref{Int_g_inf}), let's take the $\chi^{(2)}$-crystal based non-degenerate PSA as an example. The phase difference is then $\varphi = \phi_{p} - \phi_{1} - \phi_{2}$, where $\phi_{p}$ is the phase of pump, $\phi_{1}$ and $\phi_{2}$ are the phases of two input fields. In the case of $\varphi=\pi$, PSA is operating at de-amplification condition. Using the pump phase as a reference, we have $\phi_{1}=\phi_{2}=-\frac{\pi}{2}$, and the intensity of photo-current is written as
\begin{equation}\label{Int_g_inf_X-}
\begin{split}
\langle \hat i_{PSA}\rangle =  q^2 \langle \hat I_1^{out} \rangle=q^2 \frac{(G+g)^2}{16}\left[ \langle \Delta^2 \hat X_-\rangle+\langle \Delta^2  \hat Y_+\rangle \right].
\end{split}
\end{equation}
Normalizing $\langle \hat i_{PSA}\rangle$ with the corresponding SNL obtained by replacing two inputs with vacuum,
$\langle \hat i_{PSA}\rangle_{SNL}= q^2 \langle \hat I_1^{out} \rangle_{SNL}=q^2g^2$, we find the ratio
\begin{equation}
	\frac{\langle \hat i_{PSA}\rangle}{\langle \hat i_{PSA}\rangle_{SNL}}=\frac{I_s}{I_s^{SNL}}.
\end{equation}
The results indicate the intensity measured at one output of PSA can be used to characterize the inseparability of EPR source.
On the other hand, when $\varphi=0$, i.e., the PSA is operating at the amplification condition. If the entanglement relation is described by Eq. (\ref{noise_EPR}), the combination of PA and PSA with $\varphi=0$ is equivalent to a new PA with a higher gain of $G\mu-g\nu$ and $G\nu-g\mu$, which generates entangled state with noise reduction of $\langle \Delta^2 \hat X_-^{out} \rangle = \langle \Delta^2 \hat Y_+^{out} \rangle = (G-g)^2(\mu-\nu)^2$ \cite{XJ17}.

Note that the intensity $\langle \hat I_2^{out}\rangle $ measured at the other output field $\hat a^{out}_2$ is similar to $\langle \hat I_1^{out}\rangle$ in Eq. (\ref{Int_g_inf}). So the intensity measured at each output of PSA can be used to characterize the inseparability of EPR source. Moreover, our analysis shows the non-degenerate PSA followed with a power detector can be used to characterize the inseparability of entanglement as long as the phase of PSA is properly locked and the gain of PSA is high enough, while the requirement of disentangling entanglement by PSA \cite{Flurin-PRL} is not necessary.

\subsection{Pros and Cons of the new type homodyne detection}
The homodyne detections realized by PSA in Figs. \ref{fig-new-type} and \ref{fig-EPR-PA} have the advantage of detection loss tolerance in measuring the noise reduction of squeezed state and in characterizing the inseparability criterion.
For example, for the PSA with $g\rightarrow\infty$, when the detection loss exists and is modeled as $L_D$, the output intensity is decreased to $\langle \hat I{^{out}}'\rangle =(1-L_D)\langle \hat I^{out} \rangle$, and the corresponding SNL is also decreased to $\langle \hat I{^{out}}'\rangle_{SNL}=(1-L_D)\langle \hat I_{1}^{out}\rangle_{SNL}$. Hence, the normalized result of $\frac{\langle \hat I{^{out}}'\rangle}{\langle \hat I{^{out}}'\rangle_{SNL}}=\frac{\langle \hat I_1^{out}\rangle}{\langle \hat I_{1}^{out}\rangle_{SNL}}$ is irrelevant to $L_D$.
Moreover, different from the traditional method which required two BHDs, the inseparability of entanglement between two fields can be deduced from the intensity measured at one output of non-degenerate PSA (see Fig. \ref{fig-EPR-PA}). This feature brings convenience in characterizing entanglement.

However, there are bottlenecks which limit the practical application of the new type homodyne detections. First, the gain of PSA should be extremely high. Otherwise, the approximations in Eqs. (\ref{noise_degenate}) and (\ref{Int_g_inf_X-}) are not valid. In general, the achievable gain of PSA is mainly determined by the availability of pump power. The PSA with power gain higher than 20 dB ($g>10$), at which the new type homodyne are experimentally demonstrated in Refs. \cite{Shaked-NC} and \cite{Flurin-PRL}, is usually difficult realize.
Second, the noise fluctuations of $\hat X_-=\hat X_1-\hat X_2$ and $\hat Y_+ = \hat Y_1 + \hat Y_2$ of two entangled fields, which are the key of fulfilling the QIP tasks, such as quantum teleportation, quantum swapping and quantum enhanced precision measurement, can not be measured since the power detector in Figs. \ref{fig-new-type} and \ref{fig-EPR-PA} lacks the ability of resolving phase information of detected fields.
Third, the measurable bandwidth of the quantum correlation of entanglement is relatively narrow. Although the coherent combination of two entangled field realized by the nonlinear coupling in PSA is much fast than that in traditional method (see Fig. \ref{fig-tri-method}), which is realized by taking the difference and sum of the photo-currents out of two BHDs with an electronic combiner, the response bandwidth of the power detector in Fig. \ref{fig-EPR-PA} is usually very slow. From Eqs. (\ref{noise_degenate-amp}) and (\ref{HD-noise}), one sees that the amplification provided by the LO in BHD can be much higher than that by the strong pump of PSA. So the gain of the electrical amplifier $q$ in Figs. \ref{fig-new-type} and \ref{fig-EPR-PA} is usually much higher than that in Fig. \ref{fig-tri-method}(b), which severely limits the detection bandwidth.

\section{New method for measuring EPR entanglement with non-degenerate PSA assisted balanced homodyne detectors}

Having reviewed two kinds of homodyne detection and their application in characterizing EPR entanglement, we study a mew method for measuring the entanglement with a non-degenerate PSA assisted balanced homodyne detections.
During the measurement, two entangled fields are coupled into the PSA to resist the influence of detection loss before extracting the phase information of the detected fields with BHDs. The new method inherit the advantageous of the measurement methods in Secs. II and III. In this section, we will first analyze two approaches of the new method. One is to perform joint measurement by using two BHDs placed at each output of PSA. The other is to characterize the entanglement by using only one BHD at one output of PSA. Then we will study and compare the optimum operation condition for each approach by simulating the measurement results when the key parameters, such as gain of PSA and detection loss etc., are changed. At the end of this section, we will analyze the advantages and disadvantages of the new method.

\subsection{Joint measurement performed by placing two BHDs at two outputs of PSA }

\begin{figure}[htb]
	\centering
	\includegraphics[width=8cm]{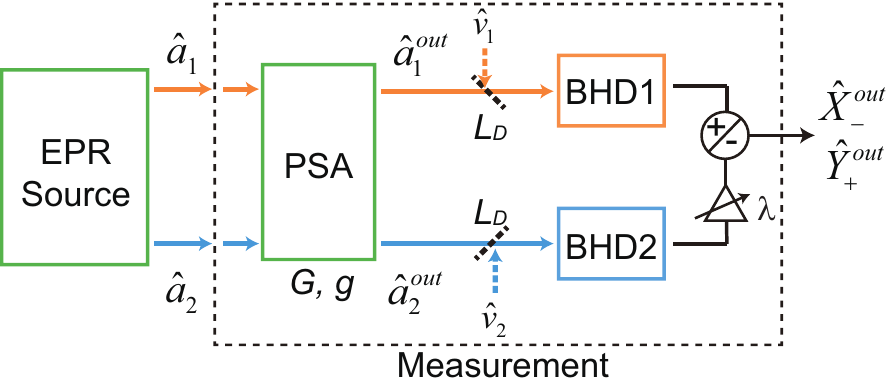}
	\caption{Entanglement measured by assisting the traditional joint measurement of two balanced homodyne detection (BHD) with a non-degenerate phase sensitive amplifier (PSA). The detection loss of each homodyne detection is modeled by a BS with reflectivity $L_D$.
	$\hat v_1/\hat v_2$ represents the vacuum field coupled into the detected fields.}
	\label{fig-PSA-joint}
\end{figure}

Figure \ref{fig-PSA-joint} shows the approach of performing joint measurement by placing two BHDs at two outputs of a PSA.
Two entangled fields $\hat a_1$ and $\hat a_2$ are coupled into the non-degenerate PSA with gain amplitude $G,g$, and the outputs $\hat a_1^{out}$ and $\hat a_2^{out}$ of PSA are respectively measured by two BHDs.
For simplicity, from here-in-after, we focus on the measured quadrature components and drop the amplification term of each BHD provided by local and electronic amplifier, i.e., $|\alpha_L|^2=1$ and $q^2=1$.
The relationship between the inputs and outputs of a PSA is same as Eq. (\ref{Gg-opa}). When PSA is operating at deamplification condition ($\varphi=\pi$), the input-output relation is:
\begin{equation}\label{Gg-opa_pi}
\begin{split}
\hat a_1^{out} &= G \hat a_1 - g \hat a_2^\dag ,\\
\hat a_2^{out} &= G \hat a_2  - g \hat a_1^\dag.
\end{split}
\end{equation}
When the LO phases of BHD1 and BHD2 are fixed at 0 or $\pi/2$, the difference or sum of photo-currents of BHD1 and BHD2 measure the quantities $\hat X_-^{out}\equiv \hat X_1^{out} - \lambda \hat X_2^{out}$ and $\hat Y_+^{out}\equiv \hat Y_1^{out} + \lambda \hat Y_2^{out}$, where $\lambda$ is the electrical gain of the current out of BHD2 to optimize the measurement results.
Hence, the noise variances of joint measurement are related to the correlation of two entangled fields, $\hat X_-$ and $\hat Y_+$, though the relation:
\begin{equation}\label{EPR-OPA-k}
\begin{split}
\langle {\Delta^2 \hat X_-^{out}} \rangle &= (G + \lambda g)^2 \langle \Delta^2 ({\hat X_1} - \frac{g + \lambda G} {G + \lambda g} {\hat X_2} ) \rangle\\
\langle {\Delta^2 \hat Y_+^{out}} \rangle &=  (G + \lambda g)^2 \langle \Delta^2 ({\hat Y_1} + \frac{g + \lambda G} {G + \lambda g} {\hat Y_2}) \rangle.
\end{split}
\end{equation}
The corresponding SNL for PSA-assisted joint measurement can be obtained by sending vacuum into both input ports of PSA. Hence, the SNL of the measurement scheme in Fig. \ref{fig-PSA-joint} is expressed as
\begin{equation}\label{I-amp-SNL2}
\begin{split}
\langle \Delta^2 \hat X_{-}^{out} \rangle_{SNL} &= \langle \Delta^2 \hat Y_{+}^{out} \rangle_{SNL} \\
&= (G + \lambda g)^2 + (g + \lambda G)^2.
\end{split}
\end{equation}
Normalizing Eq. (\ref{EPR-OPA-k}) with the SNLs, we have 
\begin{equation}\label{Noise-nor-JM}
\begin{split}
    &\langle {\Delta^2 \hat X_-^{out}} \rangle_{Nor} = \frac{\langle {\Delta^2 \hat X_-^{out}} \rangle }{\langle \Delta^2 \hat X_{-}^{out} \rangle_{SNL}}\\
        & \ = \frac{(G+\lambda g)^2}{(G + \lambda g)^2 + (g + \lambda G)^2}\langle \Delta^2 ({\hat X_1} - \frac{g + \lambda G} {G + \lambda g} {\hat X_2} ) \rangle\\
    &\langle {\Delta^2 \hat Y_+^{out}} \rangle_{Nor} = \frac{\langle {\Delta^2 \hat Y_+^{out}} \rangle }{\langle \Delta^2 \hat Y_{+}^{out} \rangle_{SNL}}\\
    & \ =\frac{(G+\lambda g)^2}{(G + \lambda g)^2 + (g + \lambda G)^2}\langle \Delta^2 ({\hat Y_1} + \frac{g + \lambda G} {G + \lambda g} {\hat Y_2} ) \rangle,
\end{split}
\end{equation}
and the inseparability 
\begin{equation}\label{I-amp-JM}
\begin{split}
I_{amp}^{JM} & = \langle {\Delta^2 \hat X_-^{out}} \rangle_{Nor} + \langle {\Delta^2 \hat Y_+^{out}} \rangle_{Nor}.
\end{split}
\end{equation}
where the superscript $JM$ stands for joint measurement.
For the case of  $\lambda = 1$ ($\frac{g + \lambda G} {G + \lambda g}=1$), we have the noise reduction due to entanglement correlation, 
\begin{equation}
\begin{split}
    \langle {\Delta^2 \hat X_-^{out}} \rangle_{Nor} &=\langle {\Delta^2 \hat X_-} \rangle_{s}=(\mu-\nu)^2,\\
    \langle {\Delta^2 \hat Y_+^{out}} \rangle_{Nor} &=\langle {\Delta^2 \hat Y_+} \rangle_{s}=(\mu-\nu)^2,
\end{split}
\end{equation}
and the inseparability in Eq. (\ref{I-amp-JM}) reach the minimum $I_s$ (see Eq. (\ref{I_s})) for arbitrary value of $g$. So the optimum results of Eqs. (\ref{Noise-nor-JM}) and (\ref{I-amp-JM}) are always the same as the same as those measured by the traditional method under perfect detection efficiency.
Moreover, the analysis indicates traditional method in Sec. IIB is just a specific case of PSA-assisted joint measurement with the pump power turning off (i.e., $g=0$).
This PSA-assisted scheme provides a parametric gain of $g>0$ for amplifying the noise correlation of entangled fields and shot noise level of the measurement with the factor $(G+\lambda g)^2$. So it is straight forward to predict that the ability of detection loss tolerance for measurement scheme in Fig. \ref{fig-PSA-joint} improves with the increase of $g$. The detailed information of how the measurement results depend on the gain of PSA and loss of detection will be given in Sec. IVC.

\subsection{Individual measurement of standard homodyne detection performed at one output of the PSA}

\begin{figure}[htb]
\centering
 \includegraphics[width=8cm]{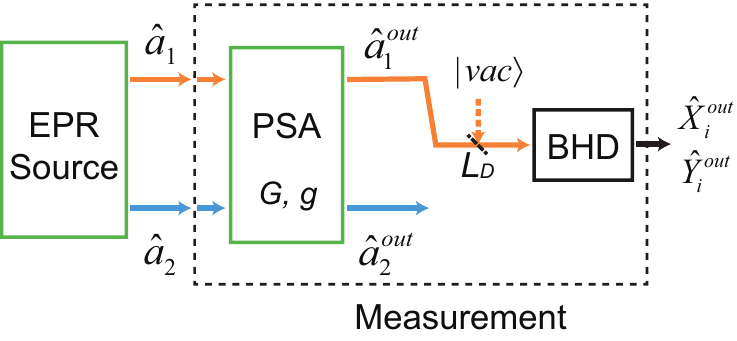}
\caption{Entanglement measurement performed by placing one balanced homodyne detection (BHD) at one output of non-degenerate phase sensitive amplifier (PSA). The detection loss of BHD is modeled by a BS with reflectivity $L_D$.}
\label{fig-PSA-ind}
\end{figure}

Indeed, as we have analyzed in Secs. IIIB and IIIC, the intensity directly measured at one output of non-degenerate PSA can be used to characterize the inseparability of entanglement. To conveniently measure the quantum correlation between the quadratures of two entangled fields at different phase angle,  we replace the power detector in Fig. \ref{fig-EPR-PA} is with a BHD, as shown in Fig. \ref{fig-PSA-ind}.

Again, we assume the non-degenerate PSA in Fig. \ref{fig-PSA-ind} is working at the de-amplification condition for measuring the entanglement produced from the scheme in Fig. \ref{fig-EPR-source}. According to
the relationship between the inputs and outputs in Eq. (\ref{Gg-opa_pi}), the variances of quadrature amplitudes $\hat X_1^{out}$ or $\hat Y_1^{out}$ at one output field, say, $\hat a_1^{out}$, are related to the correlation of the source, $\hat X_-$ and $\hat Y_+$, though the relation:
\begin{equation}\label{EPR-OPA}
	\begin{split}
		\langle \Delta^2 \hat X_1^{out} \rangle &= G^2\langle {\Delta^2 ( \hat X_1 - \frac{g}{G} \hat X_2)} \rangle, \\
		\langle \Delta^2 \hat Y_1^{out} \rangle &= G^2\langle {\Delta^2 ( \hat Y_1 + \frac{g}{G} \hat Y_2)} \rangle,
	\end{split}
\end{equation}
which are obtained by locking the LO of BHD at 0 and $\pi/2$, respectively, indicating that the noise variances of $\langle {\Delta^2 ( \hat X_1 - \frac{g}{G} \hat X_2)} \rangle$ and $\langle {\Delta^2 ( \hat Y_1 + \frac{g}{G} \hat Y_2)} \rangle$ are amplified with the factor $G^2$. On the other hand, the corresponding SNLs can be obtained by sending vacuum into both input ports of PSA. Hence, the SNLs of the measurement scheme are given by
\begin{equation}\label{I-amp-SNL}
  \langle \Delta^2 \hat X_{1}^{out} \rangle_{SNL} = \langle \Delta^2 \hat Y_{1}^{out} \rangle_{SNL}= G^2+ g^2.
\end{equation}
Normalizing Eq. (\ref{EPR-OPA}) with the SNLs in Eq. (\ref{I-amp-SNL}), we have
\begin{equation}\label{Noise-PSA-nor}
\begin{split}
    \langle {\Delta^2 \hat X_1^{out}} \rangle_{Nor} &= \frac{\langle {\Delta^2 \hat X_1^{out}} \rangle }{\langle \Delta^2 \hat X_1^{out} \rangle_{SNL}}\\
    & = \frac{G^2}{G^2+g^2}\langle \Delta^2 ({\hat X_1} - \frac{g}{G} {\hat X_2} ) \rangle,\\
    \langle {\Delta^2 \hat Y_1^{out}} \rangle_{Nor} &= \frac{\langle {\Delta^2 \hat Y_1^{out}} \rangle }{\langle \Delta^2 \hat Y_1^{out} \rangle_{SNL}}\\
    & =\frac{G^2}{G^2+g^2}\langle \Delta^2 ({\hat Y_1} + \frac{g}{G} {\hat Y_2} ) \rangle,
\end{split}
\end{equation}
and the inseparability 
\begin{equation}\label{I-OPA}
I_{amp}^{(1)}  = \langle {\Delta^2 \hat X_1^{out}} \rangle_{Nor} + \langle {\Delta^2 \hat Y_1^{out}} \rangle_{Nor},
\end{equation}
where the superscript ``$(1)$'' denotes the output field of $\hat a_1^{out}$.
When the gain of PSA is very high, i.e., $g/G \to 1$ (or $ g\rightarrow \infty$), for the EPR source in Fig. \ref{fig-EPR-source}, we have the noise reduction 
\begin{equation}
    \begin{split}
        \langle {\Delta^2 \hat X_1^{out}} \rangle_{Nor} &= \langle \Delta^2 \hat X_- \rangle_s = (\mu-\nu)^2,\\
        \langle {\Delta^2 \hat Y_1^{out}} \rangle_{Nor} &= \langle \Delta^2 \hat Y_+ \rangle_s = (\mu-\nu)^2,
    \end{split}
\end{equation}
and inseparability $I_{amp}^{(1)} \rightarrow 2(\mu-\nu)^2 $, which are exactly the same as the fluctuation and inseparability $I_s$ in Eqs. (\ref{EPR-Nor})-(\ref{I_s}). Note if the BHD is placed at the output of $\hat a^{out}_2$, the results are similar to those measured at $\hat a^{out}_1$ output, except respectively switching $g$ and $G$ in Eq. (\ref{Noise-PSA-nor}) with $G$ and $g$.

It is worth noting that the coherent combination of two entangled fields realized by non-degenerate PSA is different from that realized by subtracting or adding up the photo-currents of two BHDs, which is a linear combination of the quadrature amplitudes of $\hat X_1(\phi)$ and $\hat X_2(\phi)$. From Eq. (\ref{Gg-opa_pi}), one sees that PSA can coherently combine one input field with the conjugate of the other input field. So, in contrast to the traditional method, in which the measurement results of noise variance highly depend on LO phase of each BHD \cite{Furusawa98}, the noise variance of quadrature amplitude measured by the BHD at one output of PSA is irrelevant to LO phase \cite{JM-PRA18}. For example, for the entangled state produced by the source in Fig. \ref{fig-EPR-source}, we have the normalized noise fluctuation
\begin{equation}\label{noise-PSA1}
  \langle \Delta^2 \hat X_1^{out}(\phi) \rangle_{Nor} =  \frac{({\mu}{G} - {\nu}{g})^2 + ({\mu}{g} - {\nu}{G})^2}{(G^2+g^2)}
\end{equation}
where $\phi$ is the LO phase of BHD. Equation (\ref{noise-PSA1}) clearly indicate the variance of quadrature amplitude at arbitrary angle $\phi$, $ \langle \hat X_1^{out}(\phi) \rangle_{Nor}$, depends on the gains of both the entangled source and PSA, but does not vary with the LO phase of BHD due to the destructive quantum interference effect of entanglement in PSA \cite{Ou12}. However, to extract information carried by the entangled fields, correctly locking the phase is still a key point. For example, when one field of entanglement is encoded with the information of both weak amplitude and phase signals, we need to decoded the information by locking the phase at  0 and $\pi/2$ to respectively measure the quadrature amplitudes $\hat X_1^{out}$ and $\hat Y_1^{out}$ \cite{JM-PRA18}.

\begin{figure}[htb]
  \centering
  \includegraphics[width=8cm]{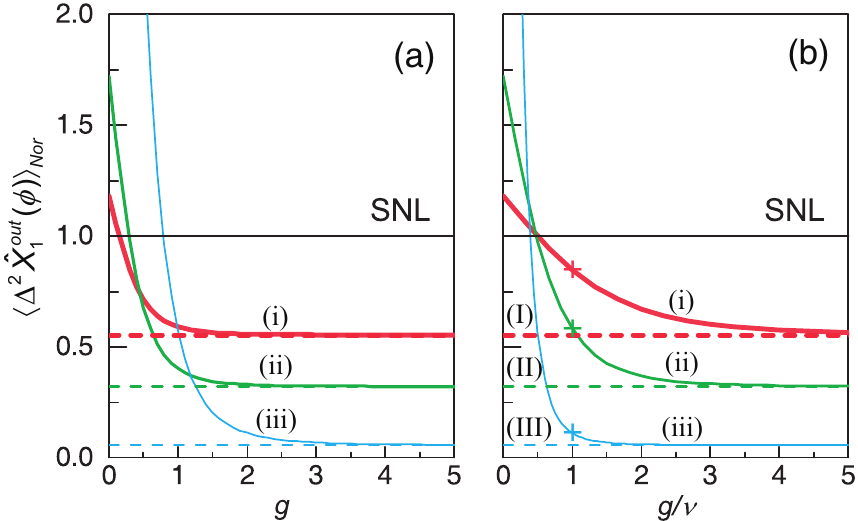}
   \caption{(a) The value of $\langle \Delta^2 \hat X_1^{out}(\phi) \rangle_{Nor}$ measured by one balanced homodyne detection versus the gain amplitude $g$ of PSA when the noise reduction of entangled source is fixed at different levels. (b) The measured value $\langle \Delta^2 \hat X_1^{out}(\phi) \rangle_{Nor}$ as a function of the ratio between the gains of PSA and source, $g/\nu$. The dashed lines labelled as (I), (II) and (III) represent the noise reduction of the source $\langle \Delta^2 \hat X_- \rangle_s = \langle \Delta^2 \hat Y_+ \rangle_s $ at the levels of $0.554, 0.321$ and $0.056$, which are obtained by setting the gain of the source $\nu$ at $0.3$, $0.6$, and $2$, respectively. The solid curves labeled as (i), (ii) and (iii) represent the measured results $\langle \Delta^2 \hat X_1^{out} (\phi)\rangle_{Nor}$ for $\langle \Delta^2 \hat X_- \rangle_{s}= \langle \Delta^2 \hat Y_+ \rangle_s =0.554, 0.321, 0.056$, respectively. The straight solid lines denote the shot noise levels (SNL).}
  \label{fig-Ent_g}
\end{figure}

Different from Eqs. (\ref{Noise-nor-JM})-(\ref{I-amp-JM}), whose optimum values are irrelevant to the gain of PSA, Eqs. (\ref{Noise-PSA-nor})-(\ref{noise-PSA1}) show that the optimized measurement of entanglement requires the satisfaction of $ g\rightarrow \infty$. In practice, it is impossible to achieve $g\rightarrow \infty$ since pump power of PSA can not be infinitely high. In order to understand how the gain of PSA affects the performance of the scheme in Fig. \ref{fig-PSA-ind}, we analyze the dependence of measurement results by taking the source in Fig. \ref{fig-EPR-source} as the example. 
Since $\langle \Delta^2 \hat X_1^{out}\rangle_{Nor} = \langle \Delta^2 \hat Y_1^{out}\rangle_{Nor} = \langle \Delta^2 \hat X_1^{out}(\phi)\rangle_{Nor} $, we will calculate $\langle \Delta^2 \hat X_1^{out}(\phi)\rangle_{Nor}$ as a function of $g$ when the noise correlation of two entangled fields directly out of the source $\langle \Delta^2 \hat X_- \rangle_s$ (or $\langle \Delta^2 \hat Y_+ \rangle_s$) is fixed at different levels, as shown in Fig. \ref{fig-Ent_g}(a) .
In the calculation, the given values $\langle \Delta^2 \hat X_- \rangle_s$ of EPR source, $0.554, 0.321$ and $0.056$ (dashed lines in Fig. \ref{fig-Ent_g}) are obtained by substituting the gain values of source $\nu=0.3, 0.6, 2$ into Eq. (\ref{EPR-Nor}); while the value $ \langle \Delta^2 \hat X_1^{out}(\phi) \rangle_{Nor}$ measured with PSA of gain $g$ is obtained by substituting the different gain values of source ($\nu$) into Eq. (\ref{noise-PSA1}).
In Fig. \ref{fig-Ent_g}, the solid curves labeled as (i), (ii) and (iii) represent the measured value $ \langle \Delta^2 \hat X_1^{out} \rangle_{Nor}$ for the EPR source with $\langle \Delta^2 \hat X_- \rangle_s$ of $0.554, 0.321$ and $0.056$, respectively. It is clear that in each case, the deviation between the measured value $\langle \Delta^2 \hat X_1^{out} (\phi) \rangle_{Nor}$ and source noise reduction level $\langle \Delta^2 \hat X_- \rangle_s$ (dashed line) decreases with the increase of $g$.  When $g$ is low, the value of $\langle \Delta^2 \hat X_1^{out} (\phi) \rangle_{Nor}$ is higher than the SNL of 1. In particular, when the pump power of PSA is zero, we have $g=0$ and $\hat X_1^{out}=\hat X_1$. In this case, the value of $\langle \Delta^2 \hat X_1^{out} (\phi) \rangle_{Nor}$ is always higher than the normalized SNL of 1 and increases with the gain of the source $\nu$ due to the thermal nature of each individual field of entanglement. When $g$ is higher than a certain level, the value of $\langle \Delta^2 \hat X_1^{out} (\phi) \rangle_{Nor}$ starts to become lower than 1. In each case, when $g$ is greater than 3, which corresponds to PSA with power gain $G^2=10$, $ \langle \Delta^2 \hat X_1^{out} \rangle_{Nor}$ becomes very close to the noise reduction of the source $\langle \Delta^2 \hat X_- \rangle_s$. In real experiment, it is easy to achieve the gain level of about $G^2=10$.

From Fig. \ref{fig-Ent_g}(a), we also find that the value of $g$ at which $\langle \Delta^2 \hat X_1^{out}(\phi)\rangle_{Nor}$ crosses the SNL of 1 increases with the decrease of $\langle \Delta^2 \hat X_- \rangle_s$. Moreover, for the EPR source with a given gain $\nu$, if we use the difference between $\langle \Delta^2 \hat X_1^{out} (\phi)\rangle_{Nor}$ and $\langle \Delta^2 \hat X_- \rangle_s$ as a standard for characterizing the reliability of measurement, it seems that in order to make the difference negligibly small, the PSA gain $g$ should increase with source gain $\nu$.
To better understand if the required gain of PSA is somehow related with the entangled degree of the EPR source, we plot $\langle \Delta^2 \hat X_1^{out}(\phi)\rangle_{Nor}$ as a function of ${g}/{\nu}$, as shown in Fig. \ref{fig-Ent_g}(b). For the given value of ${g}/{\nu}$, the difference between $\langle \Delta^2 \hat X_1^{out}(\phi) \rangle_{Nor}$ and $\langle \Delta^2 \hat X_- \rangle_s$ increase with noise level of EPR source $\langle \Delta^2 \hat X_- \rangle_s$ (dashed lines at different level). 
In particular, for the PSA function as a dis-entangler, i.e, ${g}/{\nu}=1$, at which each output of PSA is in uncorrelated vacuum state \cite{Flurin-PRL,Ou12}, we find that the measured noise of $\langle \Delta^2 \hat X_1^{out}(\phi)\rangle_{Nor}$ is far from the noise of source $\langle \Delta^2 \hat X_- \rangle_s$ when $\langle \Delta^2 \hat X_- \rangle_s$ is relatively high (or the gain of EPR source is low). Although the first demonstration of microwave entanglement was characterized by setting the PSA as a dis-entangler, the results in Fig. \ref{fig-Ent_g} indicate that for the measurement scheme in Fig. \ref{fig-PSA-ind}, the gain of PSA is the dominant factor to ensure the reliability of the measurement.

\subsection{Influence of detection loss upon the performance of the new measurement method}

Having induced the working principle of the new method, we are ready to analyze the influence of detection loss. Let's first investigate the performance of PSA assisted joint measurement scheme in Fig. \ref{fig-PSA-joint}. For the sake of convenience, we assume that the electrical gain is $\lambda=1$, and the detection losses of the two BHDs are $L_D$. With the presence of detection losses, the joint measurement quantities in Eq. (\ref{I-amp-JM}) is modified to:
\begin{equation}\label{EPR-OPA-loss-JM}
\begin{split}
\langle {\Delta^2  \hat X{_-^{out}}' } \rangle &= \langle {\Delta^2 ( \hat X{_1^{out}}' -  \hat X{_2^{out}}' )} \rangle \\
&\hskip -0.5 in = (1-L_{D})\langle {\Delta^2 ( \hat X_1^{out} -  \hat X_2^{out} )} \rangle
+2L_{D}\\
\langle {\Delta^2  \hat Y{_+^{out}}' } \rangle &=\langle {\Delta^2 ( \hat Y{_1^{out}}' +  \hat Y{_2^{out}}' )} \rangle \\
&\hskip -0.5in= (1-L_{D})\langle {\Delta^2 ( \hat Y_1^{out} +  \hat Y_2^{out} )} \rangle
+2 L_{D},
\end{split}
\end{equation}
and the corresponding SNL is
\begin{equation}\label{EPR-SNL-loss-JM}
\begin{split}
\langle {\Delta^2  \hat X{_-^{out}}' } \rangle_{SNL} &= \langle {\Delta^2  \hat Y{_+^{out}}' } \rangle _{SNL}\\
&\hskip -0.5in= 2(1-L_{D})( G +  g )^2 +2 L_{D} .
\end{split}
\end{equation}
It is straightforward to show that the second term of detection loss $2L_D$ in Eqs. (\ref{EPR-OPA-loss-JM}) and (\ref{EPR-SNL-loss-JM}) can be dropped if the gain of PSA $g$ is large. When the gain of PSA satisfies the condition
\begin{equation}\label{loss-JM}
(1-L_D)(G+ g)^2 \gg 1,
\end{equation}
the noise reduction and inseparability measured by joint measurement are modified to
\begin{equation}\label{I_Xout_loss}
\begin{split}
{\langle {\Delta^2 \hat X_-^{out}}'\rangle_{Nor}}&=\langle \Delta^2 \hat X_-  \rangle_s= (\mu - \nu)^2,\\
{\langle {\Delta^2 \hat Y_+^{out}}'\rangle_{Nor}}&=\langle \Delta^2 \hat Y_+  \rangle_s= (\mu - \nu)^2,\\
{I_{amp}^{JM}}'&=I_s,
\end{split}
\end{equation}
which means we can measure entanglement with results immune to detection loss as long as the gain of PSA high enough. For example, for the entangled fields generated by the source in Fig. \ref{fig-EPR-source}, the measured noise reduction and inseparability can be written as:
\begin{equation}
\begin{split}
    &{\langle {\Delta^2 \hat X_-^{out}}'\rangle_{Nor}}={\langle {\Delta^2 \hat Y_+^{out}}'\rangle_{Nor}} \\
    & \qquad =\frac{(1-L_D)(G+g)^2({\mu} - {\nu})^2+L_D}{(1-L_D)(G+g)^2+ L_D},
\end{split}
\end{equation}
\begin{equation}\label{I-OPA-loss4}
{I_{amp}^{JM}}' = \frac{2(1-L_D)(G+g)^2({\mu} - {\nu})^2+2L_D}{(1-L_D)(G+g)^2+ L_D}.
\end{equation}
When $(1-L_D)(G+g)^2 \gg 1$ holds, the measured values ${\langle {\Delta^2 \hat X_-^{out}}'\rangle_{Nor}}={\langle {\Delta^2 \hat Y_+^{out}}'\rangle_{Nor}}$ and ${I_{amp}^{JM}}'$ approach to $(\mu-\nu)^2$ and $2(\mu-\nu)^2$, which are the same as $\langle \Delta^2 \hat X_-  \rangle_{s}=\langle \Delta^2 \hat Y_+  \rangle_{s}$ and $I_s$  directly out of the source (see Eqs. (\ref{EPR-Nor}) and (\ref{I_s})).

\begin{figure}[htb]
	\centering
	\includegraphics[width=8.5cm]{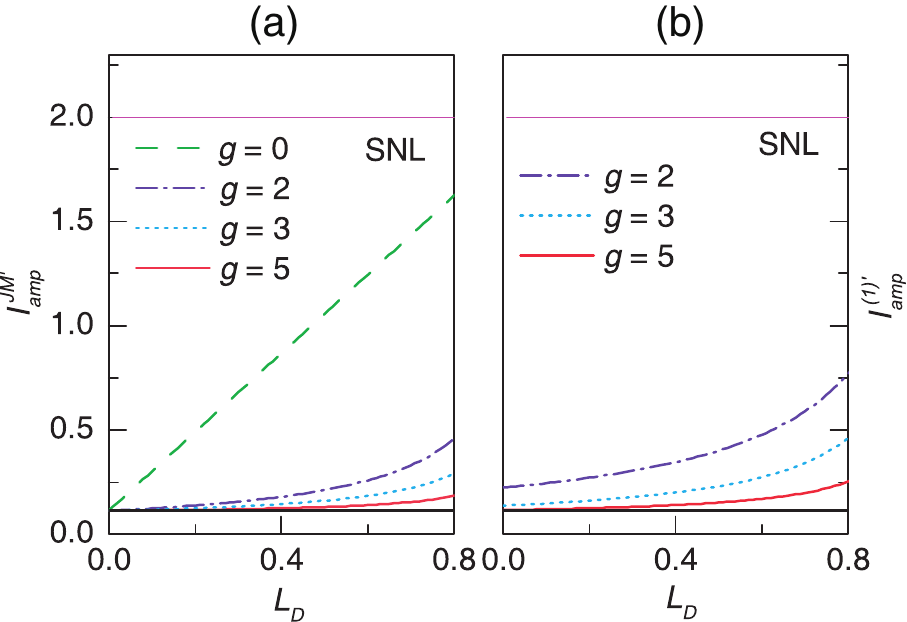}
	\caption{(a) Jointly measured inseparability ${I_{amp}^{JM}}'$ as a function the detection loss $L_D$ when the gain amplitude $g$ of PSA in Fig. \ref{fig-PSA-joint} is fixed at $g=0, 2,3,5$, respectively. (b) Inseparability ${I_{amp}^{(1)}}'$ measured by the individual BHD as a function of $L_D$ when the gain amplitude $g$ of PSA in Fig. \ref{fig-PSA-ind} is fixed at $g=2,3,5$, respectively. In the calculation, the entanglement is directly generated from the source in Fig. \ref{fig-EPR-source}, and the inseparability of the source $I_s=0.112$ represented by the thick solid lines is obtained by setting gain of source in Fig. \ref{fig-EPR-source} at $\nu=2$.		
	}
	\label{fig-loss-JM}
\end{figure}

To better illustrate influence of detection loss on the scheme in Fig. \ref{fig-PSA-joint}, we simulate the results ${I_{amp}^{JM}}'$ when the gain of PSA is fixed at different levels. The calculation is carried out by using Eq. (\ref{I-OPA-loss4}), in which the inseparability of the source $I_s=0.112$ is obtained by setting $\nu=2$  (see Fig. \ref{fig-EPR-source}). Figure \ref{fig-loss-JM}(a) plot ${I_{amp}^{JM}}'$ as a function of detection loss when the gain of PSA is respectively fixed at $g=0, 2,3, 5$. It is clear that for the case of $L_D=0$, the measured inseparability ${I_{amp}^{JM}}'$ always equal to $I_s$ and is irrelevant to gain of PSA. When $g$ is fixed, the general trend of ${I_{amp}^{JM}}'$ is to increase with $L_D$, however, the rising slope decrease with the increase $g$. In particular, for the case of $g=0$, the scheme in Fig. \ref{fig-PSA-joint} becomes equivalent to the traditional method in Fig. \ref{fig-tri-method}.
In this case, the rising slope of ${I_{amp}^{JM}}'$ versus $L_D$ is the highest. When $L_D$ increase from $0$ to $0.6$, ${I_{amp}^{JM}}'$ increases from 0.112 to 1.2. Whereas for the case of $g=5$, corresponding to PSA with power gain of about 26, the rising slope of ${I_{amp}^{JM}}'$ is very low. Even if the detection loss is $0.6$, we still have ${I_{amp}^{JM}}'= 0.13$, which is only slightly higher that  obtained with perfect detection. The results indicate the role of PSA in Fig. \ref{fig-PSA-joint} is to mitigate the influence of detection loss on entanglement measurement.

We then analyze the scheme of measuring entanglement at one output of PSA with an individual BHD when the non-ideal efficiency of BHD is taken into account. In this case, Eq. (\ref{EPR-OPA}), which describes the noise levels measured at $\hat a_1^{out}$ output of PSA-assisted scheme, has the modified form:
\begin{equation}\label{Ent-OPA-loss}
\begin{split}
     \langle {\Delta^2 \hat X_1^{out}}' \rangle  &= (1-L_{D})\langle {\Delta^2 \hat X_1^{out}  } \rangle +L_{D}\\
  &=(1-L_D)G^2  \langle {\Delta^2 ( \hat X_1 - \frac{g}{G} \hat X_2)} \rangle  +L_D,\\
     \langle {\Delta^2 \hat Y_1^{out}}'  \rangle  &= (1-L_{D})\langle \Delta^2 \hat Y_1^{out}  \rangle +L_{D}\\
     &=(1-L_D) G^2  \langle {\Delta^2 ( \hat Y_1 +  \frac{g}{G} \hat Y_2)} \rangle +L_D.
\end{split}
\end{equation}
Normalizing Eq. (\ref{Ent-OPA-loss}) with the corresponding shot noise level of $\langle {\Delta^2 \hat X_1^{out}}' \rangle_{SNL} = \langle {\Delta^2 \hat Y_1^{out}}' \rangle_{SNL} = (1-L_D)(G^2+g^2)+L_D$, Eqs. (\ref{Noise-PSA-nor}) and (\ref{I-OPA}) are modified as
\begin{equation}\label{Noise-PSA-nor-loss}
\begin{split}
    &\langle {\Delta^2 \hat X_1^{out}}' \rangle_{Nor} = \frac{(1-L_D)G^2 \langle \Delta^2 \hat X_- \rangle+L_D}{(1-L_D)(G^2+g^2)+ L_D},\\
    &\langle {\Delta^2 \hat Y_1^{out}}' \rangle_{Nor} = \frac{(1-L_D)G^2 \langle \Delta^2 \hat Y_+ \rangle+L_D}{(1-L_D)(G^2+g^2)+ L_D},\\
    &{I^{(1)}_{amp}}' = \frac{(1-L_D)G^2 [\langle \Delta^2 \hat X_- \rangle+ \langle \Delta^2 \hat Y_+ \rangle]+2L_D}{(1-L_D)(G^2+g^2)+ L_D}.
\end{split}
\end{equation}
When the relation
\begin{equation}\label{loss-one-BHD}
(1-L_D)G^2 \gg 1
\end{equation}
holds, the second term of loss $L_D$ in both numerator and denominator of Eq. (\ref{Noise-PSA-nor-loss}) can be dropped. In this case, we have
\begin{equation}\label{I-OPA-I-loss}
\begin{split}
	\langle \Delta^2  \hat X{^{out}_1}' \rangle_{Nor} &\approx \langle \Delta^2 {\hat X_-} \rangle_s,\\
	\langle \Delta^2  \hat Y{^{out}_1}' \rangle_{Nor} &\approx \langle \Delta^2 {\hat Y_+} \rangle_s,\\
	{I^{(1)}_{amp}}' &\approx I_s.
\end{split}
\end{equation}
Since the measurement obtained by placing the BHD at each output of high gain PSA is the same, the expressions in Eq. (\ref{I-OPA-I-loss}) indicate that when the gain of PSA is large enough, we can measure the entanglement of EPR source with the measurement results immune to detection loss.
Taking the entangled source in Fig. \ref{fig-EPR-source} as an example, Eq. (\ref{Noise-PSA-nor-loss}) can be rewritten as
\begin{equation}\label{I-OPA-loss3}
  \begin{split}
	&\langle \Delta^2  \hat X{^{out}_1}' \rangle_{Nor} =\langle \Delta^2  \hat Y{^{out}_1}' \rangle_{Nor}\\
	& \qquad= \frac{(1-L_D)[({\mu}{G} - {\nu}{g})^2 + ({\mu}{g} - {\nu}{G})^2]+L_D}{(1-L_D)(G^2+g^2)+ L_D},\\
    &{I_{amp}^{(1)}}' = \frac{2(1-L_D)[({\mu}{G} - {\nu}{g})^2 + ({\mu}{g} - {\nu}{G})^2]+2L_D}{(1-L_D)(G^2+g^2)+ L_D}.
  \end{split}
\end{equation}
When $(1-L_D)G^2 \gg 1$ holds, the measured inseparability ${I_{amp}^{(1)}}'$ approaches to $2(\mu-\nu)^2$, which is the ideal measurement result of $I_s$ (see Eq. (\ref{I_s})).

We have shown in Sec. IVB that measuring entanglement with one BHD only works for the PSA with high gain. So we will only analyze the influence of loss on the measurement when the gain of PSA $g$ in Fig. \ref{fig-PSA-ind} is relatively high. We carry out the calculation by using Eq. (\ref{I-OPA-loss4}), in which the entangled source with $I_s=0.112$ is the same as in Fig. \ref{fig-loss-JM}(a). Figure \ref{fig-loss-JM}(b) shows $ {I_{amp}^{(1)}}' $ as a function of detection loss when the gain of PSA is respectively fixed at $g=2, 3, 5$.
One sees that in each case, $ {I_{amp}^{(1)}}' $ slightly increases with $L_D$. Moreover, the rising slope of $ {I_{amp}^{(1)}}' $ decreases with the increase of $g$. 
Note that even if the detection efficiency if perfect ($L_D=0$), $ {I_{amp}^{(1)}}' $ slightly deviate from the thick solid line, representing the deviation between ${I_{amp}^{(1)}}'$ and $I_s$ decrease with the increase $g$. The results indicate that in order to measure entanglement with only one BHD, the gain of PSA in Fig. \ref{fig-PSA-ind} should be high enough even if the detection efficiency is high.
Moreover, comparing Fig. \ref{fig-loss-JM}(b) with Fig. \ref{fig-loss-JM}(a), we find that if the gain of PSA is the same, the deviation between ${I_{amp}^{JM}}'$ and $I_s$ is always smaller than that between ${I_{amp}^{(1)}}'$ and $I_s$. So the loss tolerance feature for the approach in Fig. \ref{fig-PSA-joint} is better than that in Fig. \ref{fig-PSA-ind}.

\subsection{Pros and cons of the new measurement method}

The results in Fig. \ref{fig-loss-JM} clearly shows that comparing with the traditional method (see the curve for $g=0$ (or $G=1$) in Fig. \ref{fig-loss-JM}(a)), both approaches of the new method have the advantages of detection loss tolerance. If the measurement results are obtained from the approach of joint measurement of two BHDs, the ability of loss tolerance increases with $g$, and the new method is always better than the traditional one as long as pump power of PSA is non-zero. If only one output of PSA is measured by a BHD, the advantages of new method can be exhibited only when the gain (or pump power) of PSA is high enough.
Hence, the scheme in Fig. \ref{fig-PSA-ind} seems to be less attractive.  However, an outstanding feature for the approach in Fig. \ref{fig-PSA-ind} is that only one BHD is required, which is extremely useful for characterizing or measuring entanglement between two different types of waves. For example, for the entanglement between atom and light, the quantum correlation between two field can be characterized by only performing homodyne detection on the optical field.

Another potential advantage provided by the new method is that the response bandwidth of detection can be tremendously improved. On the one hand, the coherent combination of two entangled fields realized by the nonlinear coupling in high gain PSA is much fast than that the electrical combiner used in traditional method. On the other hand, the noises of entangled state measured by the new method have experienced two kind of amplification. In addition to the amplification provided by the LO of BHD, the noises are amplified by PSA under strong pump. As shown in Eqs. (\ref{SNL-PSA}), (\ref{I-amp-SNL2}) and (\ref{I-amp-SNL}), the SNL of the new measurement method is lifted by the high gain PSA. Therefore, to achieve of the goal of effectively amplifying a weak input signal to a level that is otherwise buried in the classical noise such as detector's dark current, thermal electronic noise of current amplifiers and ambient background light, gain of electronic amplifier can be relaxed, which  will result in a increased gain bandwidth in BHD.

Despite these advantages, it is worth noting that the new method can only overcome the problem of detection loss. For the losses occurred in coupling the EPR state into the PSA, on the other hand, the usage of PSA is not useful because these losses have negative effects on the entangled degree, which is equivalent to the detection loss for the traditional method, (see Eq. (\ref{I-loss}) in Sec. IIC). 
Therefore, it is crucial to ensure that the coupling efficiency between the entangled source and PSA is as high as possible.

\section{Measurement of multi-mode entangled states}

In previous sections, each field of entanglement is viewed as in single mode. In practice, we always have to deal with systems in multiple modes. Particularly, for the entanglement generated by pulse excited system, each field of entanglement is composed of multiple frequency components. In this section we will extend the models of EPR source and measurement schemes from single-mode to multi-mode. After briefly describing the problems of measuring multi-mode entanglement by using the traditional method depicted in Fig. \ref{fig-tri-method}, we will analyze the performance of the new method and show its advantages in measuring multi-temporal mode entangled state.

\subsection{Entanglement generated from pulse pumped parametric amplifier}

If the entangled source in Fig. \ref{fig-EPR-source} is realized by pumping the PA with a train of laser pulses, the two output fields will exhibit multi-mode nature \cite{Wasilewski06}. However, no matter how complicated the system is, a parametric amplification, when treated as a linear device, can always be viewed as superposition of its eigen-modes which do not change after the amplifier. Assume the nonlinear medium of parametric amplifier is in waveguide structure and only supports single spatial mode, there exists an independent set of pairwise temporal modes $\{\hat A_j,\hat B_j\}~(j=1,2,...)$ for the entangled fields, and the input-output relation can be expressed as \cite{sil,guo15}:
\begin{equation}\label{AB}
\begin{split}
   \hat A_j &= \mu_j \hat A_j^{in} + \nu_j \hat B_j^{{in}\dag},\\
   \hat B_j &= \mu_j \hat B_j^{in} + \nu_j \hat A_j^{{in}\dag},
\end{split}
\end{equation}
where
\begin{equation}\label{A-B}
\begin{split}
   \hat A_j^{\dag} \equiv \int d\omega_1 \phi_j(\omega_1) \hat a_1^{\dag}(\omega_1) ,\\
   \hat B_j^{\dag} \equiv \int d\omega_2 \psi_j(\omega_2) \hat a_2^{\dag}(\omega_2)
\end{split}
\end{equation}
define the creation operators of two entangled fields $\hat a_1$ and $\hat a_2$, and $\mu_j$, $\nu_j$ with $\mu_j^2-\nu_j^2 =1$ ($\mu_1 \geqslant \mu_2 \geqslant \mu_3 \cdots$) are the gain coefficient for $j-$th mode. The complex functions $\phi_j(\omega_1)$ and $\psi_j(\omega_2)$  satisfying the orthogonal relations, $\int d\omega_1 \phi_i^*(\omega_1)\phi_j(\omega_1) =\delta_{ij}=\int d\omega_2 \psi_i^*(\omega_2)\psi_j(\omega_2)$, represent the spectrum of two entangled fields in the $j-$th order temporal modes  $f_j(\tau)= \int d\omega_s \phi_j(\omega_s) e^{-i \omega_s \tau}$ and $h_j(\tau)= \int d\omega_i \psi_j(\omega_i) e^{-i \omega_i \tau}$.

The noise variances for the difference and sum of quadrature amplitude of $j-$th mode entanglement are
\begin{equation}\label{noise_EPR_multi}
\begin{split}
	\langle \Delta^2 \hat X_- \rangle_j &=  \langle \Delta^2 (\hat X_{1j} -\hat X_{2j}) \rangle =2 (\mu_j-\nu_j)^2,\\
	\langle \Delta^2 \hat Y_+ \rangle_j &=  \langle \Delta^2 (\hat Y_{1j} +\hat Y_{2j}) \rangle =2 (\mu_j-\nu_j)^2,
 \end{split}
 \end{equation}
where $\hat X_j = \hat A_j + \hat A_j^\dag$, $\hat Y_j = -i (\hat A_j -\hat A_j^\dag)$ are the quadrature-phase amplitudes for $j$-th modes.
The corresponding SNL for each mode 
\begin{equation} \label{SNL_EPR_multi}
    \langle \Delta^2 \hat X_- \rangle_{j-SNL} = \langle \Delta^2 \hat Y_+ \rangle_{j-SNL} =2
\end{equation}
can be obtained by replacing the output fields $\hat a_1, \hat a_2$ with vacuum. 



For the two entangled fields in $j$-th mode pair, $\phi_j(\omega_1)$ and $\psi_j(\omega_2)$, the normalized noise variances and inseparability of the source are written as:
\begin{equation}
    \begin{split}
        \langle \Delta^2 \hat X_- \rangle_{j-s}&=\frac{\langle \Delta^2 \hat X_- \rangle_{j}}{\langle \Delta^2 \hat X_- \rangle_{j-SNL}}=(\mu_j-\nu_j)^2\\
        \langle \Delta^2 \hat Y_+ \rangle_{j-s}&=\frac{\langle \Delta^2 \hat Y_+ \rangle_{j}}{\langle \Delta^2 \hat Y_+ \rangle_{j-SNL}}(\mu_j-\nu_j)^2
    \end{split}
\end{equation}
\begin{equation}
    {\rm and} ~~ I_{j-s}=\langle \Delta^2 \hat X_- \rangle_{j-s}+\langle \Delta^2 \hat Y_+ \rangle_{j-s} = 2(\mu_j-\nu_j)^2.
\end{equation}

\subsection{Measuring pulsed entanglement with the traditional method}

When the entanglement between two pulsed fields is measured by the traditional method in Fig. \ref{fig-tri-method}, the fields of LOs for two BHDs, $A_{LO}(t), B_{LO}(t)$, should be in pulsed form as well. If the two LOs are transform-limited pulses, we have
\begin{equation}\label{A-B-LO}
\begin{split}
   A_{LO}(t) = {\cal E}e^{i\phi_0} \frac{1}{\sqrt{2\pi}} \int d\omega_1  A_{LO}(\omega_1) e^{-i\omega_1 t},\\
   B_{LO}(t) = {\cal E}e^{i\psi_0} \frac{1}{\sqrt{2\pi}} \int d\omega_2  B_{LO}(\omega_2) e^{-i\omega_2 t},
\end{split}
\end{equation}
where the fields $A_{LO}(\omega_1),B_{LO}(\omega_2)$ satisfy the normalization condition $\int d\omega_1  |A_{LO}(\omega_1)|^2 =1$ and $ \int d\omega_2  |B_{LO}(\omega_2)|^2 =1$, $\phi_0,\psi_0$ are the global phases of two LOs, and ${\cal E}$ denotes amplitude of each LO.
When the amplitude of each LO is much stronger than the detected fields, i.e., ${\cal E}\gg 1$, the output currents of two BHDs are given by
\begin{equation}\label{i1i2}
\begin{split}
   \hat i_1 = \int dt [A^*_{LO}(t) \hat E_1(t) + h.c.],\\
   \hat i_2 = \int dt [B^*_{LO}(t) \hat E_2(t) + h.c.],
\end{split}
\end{equation}
where $\hat E_{1,2}(t)= \frac{1}{\sqrt{2\pi}}\int d\omega \hat a_{1,2}(\omega)e^{-i\omega t}$ are the field operators for the detected fields.

With the orthogonal modes defined in Eq. (\ref{A-B}), the LO fields can be decomposed as
\begin{equation}\label{A-B-de}
\begin{split}
   A_{LO}(\omega) = \sum_j \xi_j\phi_j(\omega),\\
   B_{LO}(\omega) = \sum_j \zeta_j\psi_j(\omega),
\end{split}
\end{equation}
where
\begin{equation}\label{xi-zt}
\begin{split}
   \xi_j\equiv |\xi_j|e^{i\theta_j} = \int d\omega A_{LO}(\omega)\phi_j^*(\omega),\\
   \zeta_j\equiv |\zeta_k|e^{i\theta'_j} = \int d\omega B_{LO}(\omega)\psi_j^*(\omega)
\end{split}
\end{equation}
with $\sum_j |\xi_j|^2=1=\sum_j |\zeta_j|^2$ are the complex coefficients characterizing the mode matching. $\theta_j, \theta_j'$ are the LO phases for measuring the two entangled fields in $j-$th mode. Accordingly, the output currents in Eq. (\ref{i1i2}) can be rewritten as
\begin{equation}\label{i1i2-2}
\begin{split}
   \hat i_1 = {\cal E}\sum_j|\xi_j| \hat X_{1j}(\theta_j+\phi_0),\\
    \hat i_2 = {\cal E}\sum_j|\zeta_j| \hat X_{2j}(\theta'_j+\psi_0),
\end{split}
\end{equation}
where $\hat X_{1j}(\theta_j)\equiv \hat A_j e^{-i\theta_j}+\hat A_j^{\dag} e^{i\theta_j}$, $\hat X_{2j}(\theta'_j)\equiv \hat B_j e^{-i\theta'_j}+\hat B_j^{\dag} e^{i\theta'_j}$ are the $j$-th quadrature-phase amplitudes for the modes.
To obtain the quantum correlation between the quadrature amplitudes of two fields, we measure variance in current difference and sum
\begin{equation}\label{Di1i2}
\begin{split}
\langle \Delta^2(\hat i_1 \mp\hat i_2)\rangle &= {\cal E}^2 \sum_j \left\langle \Delta^2[|\xi_j| \hat X_{1j}(\theta_j+\phi_0) \right.\\
 & \quad \left. \mp |\zeta_j| \hat X_{2j}(\theta_j'+\psi_0)]\right\rangle.
\end{split}
\end{equation}

It is well known that the entanglement degree, reflected by the inseparability, is defined through a definite quadrature-phase amplitude and its
orthogonal component of two fields, and the measurement output depends on different quadrature-phase angles. So we set the LO phases at two sets of orthogonal angles: $\phi_0,\psi_0$ and $\phi_0+\pi/2, \psi_0-\pi/2$. To optimize the measurement of entanglement, We then investigate how to minimize the calculated inseparability by changing $\phi_0,\psi_0$.

For the entangled fields described in Eq. (\ref{AB}), we calculate the inseparability from traditional method through the quantity in Eq. (\ref{Di1i2}) and the definition of inseparability in Eq. (\ref{I_s}) leads to
\begin{equation}\label{I-multi}
  I^{multi} = \frac{1}{2}\sum_j I_j
\end{equation}
with $I_j \equiv  \langle \Delta^2[|\xi_j|\hat X_{1j}(\theta_j+\phi_0) - |\zeta_j|\hat X_{2j}(\theta_j'+\psi_0) ] \rangle + \langle \Delta^2[|\xi_j|\hat X_{1j}(\theta_j+\phi_0+\pi/2) - |\zeta_j|\hat X_{2j}(\theta_j'+\psi_0-\pi/2)]\rangle$.
Note that for single mode case, we have $\xi_j=\delta_{j,j_0}=\zeta_j$ and we can recover the expression for inseparability in Eq. (\ref{I_s}).
For the multi-mode entangled source described by Eq. (\ref{AB}), we find
\begin{equation}\label{I-j}
\begin{split}
      I_j &= 2\big|\mu_j \xi_j - \nu_j \zeta_j^* e^{-i(\phi_0+\psi_0)}\big|^2 + 2\big|\nu_j \xi_j - \mu_j \zeta_j^* e^{-i(\phi_0+\psi_0)}\big|^2 \\
  &=2(\mu_j^2+\nu^2_j)(|\xi_j|^2+|\zeta_j|^2)-8\mu_j\nu_j|\xi_j||\zeta_j|\cos\Delta\theta_j
\end{split}
\end{equation}
with $\Delta\theta_j\equiv \theta_j+\theta'_j+\phi_0+\psi_0$. Although $\theta_j$ and $\theta'_{j}$ vary with $j$, the sum $\theta_j+\theta'_j$ inherit the phase of pump and does not change with $j$. So Eq. (\ref{I-j}) indicates that the nonsynchronous phase of twin beams in different mode pairs will not influence the measured result of $I_j$.
If the LOs match one specific pair of modes of the entangled fields, labelled as $j_0$-th mode, we have $\xi_j=\delta_{jj_0}=\zeta_j$ and $I^{multi} = 2(\mu_{j_0}-\nu_{j_0})^2$. This is the same as the single mode case discussed earlier. Otherwise, the inseparability obtained from homodyne detection is
\begin{equation}\label{I-multi-2}
  I^{multi} = \sum_j \big[ (\mu_j |\xi_j| - \nu_j |\zeta_j|)^2 + (\nu_j |\xi_j| - \mu_j |\zeta_j|)^2\big],
\end{equation}
which can be viewed as an averaged value of inseparability of each mode pair. For the pulsed pump with Gaussian shaped spectrum, the gain
coefficient of the fundamental mode $\nu_j$ with $j=1$ is the highest \cite{LNN16}. If we can perfectly match the modes of LOs with the two entangled fields in fundamental modes, we will be able to obtain the lowest value of inseparability coefficient, $I_1$, i.e., $I^{multi}<I_1$. However, the method of precisely knowing the temporal-mode profile has not be available yet, so it is difficult to directly measure $I_1$ by properly shaping the spectrum of LOs. 
Moreover, when the mode matching coefficients for the two entangled fields $\xi_j$ and $\zeta_j$ are significantly different from each other, the measured value of $I^{multi}$ might become higher than 2 due to the thermal nature of individual field. 
For example, assuming the entanglement is a super position of two pairs of temporal
modes, if the mode matching coefficients between the detected fields and local oscillators are $|\xi_1|=0,~|\xi_2|=1$ and $|\zeta_1|=1,~|\zeta_2|=0$, the measured value of $I^{multi}= \mu_1^2+\mu_2^2+\nu_1^2+\nu_2^2$ is higher than 2, and the inseparability criterion can not be obtained.
So the entangled degree of pulsed entanglement measured by traditional method is always smaller than what is anticipated.
Next, we will analyze the performance of the new method in measuring pulsed entanglement.

\subsection{Measurement using the PSA assisted balanced homodyne detections}

To clearly illustrate the advantageous of new method in multi-mode case, we first take the PSA-assisted scheme in Fig. \ref{fig-PSA-ind} as an example. Assuming the PSA has the same spectrum property as the entangled source but has a different set of gains: $G_j,g_j$. From mode decomposition of the pulse-pumped parametric proess \cite{sil,guo15},
we know that $G_j = \cosh(r_j G'),g_j=\sinh(r_j G')$ with $r_1> r_2> ...$ as the mode number and $\sum r_j^2=1$, where $G'$ is the gain parameter related to the pump power.
The input-output relation for the PSA is similar to Eq. (\ref{AB}) but with $\mu_j,\nu_j$ replaced by the PSA's gain parameters $G_j,-g_j$. Here, we choose $-g_j$ to describe the PSA in de-amplification.

Assuming the PSA gain is large ($G_j\approx g_j \gg 1$), we find the measured fluctuations of the quadrature-phase amplitudes of each mode at two outputs of the amplifier can be respectively written as
\begin{equation}\label{X1X2-amp}
\begin{split}
   \langle \Delta^2 \hat X_{1j}^{out}(\theta_j+\phi_0)\rangle &= 2|\xi_j|^2 G_j^2(\mu_j-\nu_j)^2\\
   \langle \Delta^2 \hat X_{2j}^{out}(\theta'_j+\psi_0)\rangle &= 2|\zeta_j|^2G_j^2(\mu_j-\nu_j)^2.
\end{split}
\end{equation}
The SNL measured at $\hat a_1^{out}$ is $\langle \Delta^2 \hat X_{1j}^{out}(\theta_j+\phi_0)\rangle_{SNL} = 2\sum_j |\xi_j|^2 G_j^2$, 
so the normalized values $\langle \Delta^2 \hat X_{1j}^{out}\rangle_{Nor} $ and $\langle \Delta^2 \hat Y_{1j}^{out}\rangle_{Nor}$  are
\begin{equation}
    \langle \Delta^2 \hat X_{1j}^{out}\rangle_{Nor} = \langle \Delta^2 \hat Y_{1j}^{out}\rangle_{Nor}= \frac{\sum_j |\xi_j|^2 G_j^2(\mu_j-\nu_j)^2}{\sum_j |\xi_j|^2G_j^2}.
\end{equation}
As a result, the inseparability measured from output port $\hat a_1^{out}$ is
\begin{equation}\label{I-amp-multi}
I^{multi}_{amp} = \frac{2\sum_j|\xi_j|^2G_j^2(\mu_j-\nu_j)^2}{\sum_j |\xi_j|^2 G_j^2}.
\end{equation}
For the output $\hat a_2^{out}$ of PSA, the measured inseparability is similar as Eq. (\ref{I-amp-multi}) but replacing $\xi_j$ with $\zeta_j$.
Since $r_1> r_2>...$,  we have $G_1 = \cosh (r_1G')\gg \cosh (r_2G')=G_2 \gg G_3 ...$ For large $G'$, Eq. (\ref{I-amp-multi}) can be approximated as
\begin{equation}\label{I-amp2}
I^{multi}_{amp} \approx \frac{2|\xi_1|^2G_1^2(\mu_1-\nu_1)^2}{|\xi_1|^2 G_1^2} =2(\mu_1-\nu_1)^2
\end{equation}
for $G'\rightarrow \infty$.

Similarly, for the PSA-assisted joint measurement in Fig. \ref{fig-PSA-joint}, the inseparability is
\begin{equation}\label{I-amp-JM-multi}
I^{multi}_{amp-JM} = \frac{2\sum_j (|\xi_j|+|\zeta_j|)^2G_j^2(\mu_j-\nu_j)^2}{\sum_j (|\xi_j|+|\zeta_j|)^2 G_j^2}
\end{equation}
for PSA with large gain. Since $r_1> r_2>...$, Eq. (\ref{I-amp-JM-multi}) can be approximated as
\begin{equation}\label{I-amp-JM2}
I_{multi}^{amp-JM} \approx \frac{2(|\xi_1|+|\zeta_1|)^2G_1^2(\mu_1-\nu_1)^2}{|(|\xi_1|+|\zeta_1|)^2 G_1^2} =2 (\mu_1-\nu_1)^2
\end{equation}
for $G'\rightarrow \infty$.
Comparing Eqs. (\ref{I-amp-multi}) and (\ref{I-amp-JM-multi}) with Eq. (\ref{I-multi-2}), one sees that the high gain PSA can selectively amplify the entangled fields in mode pair with $j=1$. By properly setting the LO phase of BHD, the variances $\langle \Delta^2 \hat X_- \rangle_{1-s}$, $\langle \Delta^2 \hat Y_+ \rangle_{1-s}$ and inseparability $I_{1-s}$ can be measured, and the results are immure to detection loss \cite{LJM19}.
Therefore, the high gain PSA assisted BHD is better than that measured by using traditional method. So the new measurement method is advantageous over the traditional method in the multi-mode case.

\section{Summary and discussion}

In summary, we have investigated a new method of detecting continuous variable entanglement by using phase sensitive amplifier assisted balanced homodyne detections. The new method is robust to a large number of key issues in the detection process of quantum state, namely non-unit quantum efficiency detectors and mode-mismatch between the detected multi-mode fields and the local oscillators. Additionally, the new method allows to measure quantum entanglement even when one of the fields is not optically accessible, which effectively simplifying the experimental resources required by the traditional balanced homodyne measurements. Moreover, there exists the potential of tremendously increasing bandwidth in measuring the quantum correlation. The new method is applicable to many relevant problems in quantum optics.

The new method for measuring quantum entanglement should also be beneficial in areas such as quantum information and quantum metrology where entangled quantum states are used for performance enhancement.
Although we have only discussed the noise measurement, it works equally well when signals are considered. Indeed, the underlying physics in a few experiments of using SU(1,1) nonlinear interferometer for quantum metrology is that the advantage of entangled sources is maintained for the enhancement of the signal-to-noise ratio even in the presence of losses at detection \cite{liu-OE19,sil,guo15}, which is the same as the new method.

In this paper, we have studied the new measurement method by taking two entangled fields with positive correlation between $\hat X_1$ and $\hat X_2$ and with negative correlation between $\hat Y_1$ and $\hat Y_2$ as the example.
In addition, we assume the correlation between two entangled fields is symmetric. We believe the new method is suitable for measuring various kind of entanglement.
For example, another type of entanglement with negative correlation between $\hat X_1$ and $\hat X_2$ and with pissitive correlation between $\hat Y_1$ and $\hat Y_2$ \cite{li02} can also be measured by the new method, but the PSA need to be operated at the amplification condition, at which the noise variance at each output of PSA is proportional to product between the PSA gain $G^2$ and the variances of operators $\hat X_1+\hat X_2$ (or $\hat Y_1-\hat Y_2$).
Moreover, for the entanglement with asymmetric correlation between two fields, i.e., $\hat X_1 \mp k{\hat X_2}$ and $\hat Y_1 \pm k{\hat Y_2}$ ($k\neq 1$), we can optimize the joint measurement result by accordingly changing the electrical gain to adjust the photo-current out of one BHD. If only one BHD is placed at one output of PSA, the optimized measurement can be obtained by properly changing the gain PSA.

Finally, it is worth pointing out that the new method of assisting BHD with high gain PSA-assisted BHD is useful in the measuring the noise reduction and inseparability of entanglement. However, the PSA, which functions as a new type homodyne detection for realizing the measurement, can not be viewed as a tool for quantum state transformation. This is because the noise variance for the difference and sum of the quadrature amplitudes of the PSA output fields are not lower than vacuum noise, especially for the PSA with high gain, as illustrated by  Eq. (\ref{EPR-OPA-k}).

\section*{ACKNOWLEDGMENTS}
The work is supported in part by the National Natural Science Foundation of China (Grants No. 91836302, No. 91736105, and No. 11527808).

\nocite{*}


\begin{thebibliography}{35}%
	\makeatletter
	\providecommand \@ifxundefined [1]{%
		\@ifx{#1\undefined}
	}%
	\providecommand \@ifnum [1]{%
		\ifnum #1\expandafter \@firstoftwo
		\else \expandafter \@secondoftwo
		\fi
	}%
	\providecommand \@ifx [1]{%
		\ifx #1\expandafter \@firstoftwo
		\else \expandafter \@secondoftwo
		\fi
	}%
	\providecommand \natexlab [1]{#1}%
	\providecommand \enquote  [1]{``#1''}%
	\providecommand \bibnamefont  [1]{#1}%
	\providecommand \bibfnamefont [1]{#1}%
	\providecommand \citenamefont [1]{#1}%
	\providecommand \href@noop [0]{\@secondoftwo}%
	\providecommand \href [0]{\begingroup \@sanitize@url \@href}%
	\providecommand \@href[1]{\@@startlink{#1}\@@href}%
	\providecommand \@@href[1]{\endgroup#1\@@endlink}%
	\providecommand \@sanitize@url [0]{\catcode `\\12\catcode `\$12\catcode
		`\&12\catcode `\#12\catcode `\^12\catcode `\_12\catcode `\%12\relax}%
	\providecommand \@@startlink[1]{}%
	\providecommand \@@endlink[0]{}%
	\providecommand \url  [0]{\begingroup\@sanitize@url \@url }%
	\providecommand \@url [1]{\endgroup\@href {#1}{\urlprefix }}%
	\providecommand \urlprefix  [0]{URL }%
	\providecommand \Eprint [0]{\href }%
	\providecommand \doibase [0]{https://doi.org/}%
	\providecommand \selectlanguage [0]{\@gobble}%
	\providecommand \bibinfo  [0]{\@secondoftwo}%
	\providecommand \bibfield  [0]{\@secondoftwo}%
	\providecommand \translation [1]{[#1]}%
	\providecommand \BibitemOpen [0]{}%
	\providecommand \bibitemStop [0]{}%
	\providecommand \bibitemNoStop [0]{.\EOS\space}%
	\providecommand \EOS [0]{\spacefactor3000\relax}%
	\providecommand \BibitemShut  [1]{\csname bibitem#1\endcsname}%
	\let\auto@bib@innerbib\@empty
	\bibitem [{\citenamefont {Braunstein}\ and\ \citenamefont
		{Kimble}(2000)}]{bra00}%
	\BibitemOpen
	\bibfield  {author} {\bibinfo {author} {\bibfnamefont {S.~L.}\ \bibnamefont
			{Braunstein}}\ and\ \bibinfo {author} {\bibfnamefont {H.~J.}\ \bibnamefont
			{Kimble}},\ }\bibfield  {title} {\bibinfo {title} {Dense coding for
			continuous variables},\ }\href {https://doi.org/10.1103/PhysRevA.61.042302}
	{\bibfield  {journal} {\bibinfo  {journal} {Phys. Rev. A}\ }\textbf {\bibinfo
			{volume} {61}},\ \bibinfo {pages} {042302} (\bibinfo {year}
		{2000})}\BibitemShut {NoStop}%
	\bibitem [{\citenamefont {Zhang}\ and\ \citenamefont {Peng}(2000)}]{zh00}%
	\BibitemOpen
	\bibfield  {author} {\bibinfo {author} {\bibfnamefont {J.}~\bibnamefont
			{Zhang}}\ and\ \bibinfo {author} {\bibfnamefont {K.}~\bibnamefont {Peng}},\
	}\bibfield  {title} {\bibinfo {title} {Quantum teleportation and dense coding
			by means of bright amplitude-squeezed light and direct measurement of a bell
			state},\ }\href@noop {} {\bibfield  {journal} {\bibinfo  {journal} {Phys.
				Rev. A}\ }\textbf {\bibinfo {volume} {62}} (\bibinfo {year}
		{2000})}\BibitemShut {NoStop}%
	\bibitem [{\citenamefont {Li}\ \emph {et~al.}(2002)\citenamefont {Li},
		\citenamefont {Pan}, \citenamefont {Jing}, \citenamefont {Zhang},
		\citenamefont {Xie},\ and\ \citenamefont {Peng}}]{li02}%
	\BibitemOpen
	\bibfield  {author} {\bibinfo {author} {\bibfnamefont {X.}~\bibnamefont
			{Li}}, \bibinfo {author} {\bibfnamefont {Q.}~\bibnamefont {Pan}}, \bibinfo
		{author} {\bibfnamefont {J.}~\bibnamefont {Jing}}, \bibinfo {author}
		{\bibfnamefont {J.}~\bibnamefont {Zhang}}, \bibinfo {author} {\bibfnamefont
			{C.}~\bibnamefont {Xie}},\ and\ \bibinfo {author} {\bibfnamefont
			{K.}~\bibnamefont {Peng}},\ }\bibfield  {title} {\bibinfo {title} {Quantum
			dense coding exploiting a bright {Einstein-Podolsky-Rosen} beam},\
	}\href@noop {} {\bibfield  {journal} {\bibinfo  {journal} {Phys. Rev. Lett.}\
		}\textbf {\bibinfo {volume} {88}},\ \bibinfo {pages} {047904} (\bibinfo
		{year} {2002})}\BibitemShut {NoStop}%
	\bibitem [{\citenamefont {Steinlechner}\ \emph {et~al.}(2013)\citenamefont
		{Steinlechner}, \citenamefont {Bauchrowitz}, \citenamefont {Meinders},
		\citenamefont {Muller-Ebhardt}, \citenamefont {Danzmann},\ and\ \citenamefont
		{Schnabel}}]{snb13}%
	\BibitemOpen
	\bibfield  {author} {\bibinfo {author} {\bibfnamefont {S.}~\bibnamefont
			{Steinlechner}}, \bibinfo {author} {\bibfnamefont {J.}~\bibnamefont
			{Bauchrowitz}}, \bibinfo {author} {\bibfnamefont {M.}~\bibnamefont
			{Meinders}}, \bibinfo {author} {\bibfnamefont {H.}~\bibnamefont
			{Muller-Ebhardt}}, \bibinfo {author} {\bibfnamefont {K.}~\bibnamefont
			{Danzmann}},\ and\ \bibinfo {author} {\bibfnamefont {R.}~\bibnamefont
			{Schnabel}},\ }\bibfield  {title} {\bibinfo {title} {Quantum-dense
			metrology},\ }\href@noop {} {\bibfield  {journal} {\bibinfo  {journal} {Nat
				Photon}\ }\textbf {\bibinfo {volume} {7}},\ \bibinfo {pages} {626} (\bibinfo
		{year} {2013})}\BibitemShut {NoStop}%
	\bibitem [{\citenamefont {Bennett}\ \emph {et~al.}(1993)\citenamefont
		{Bennett}, \citenamefont {Brassard}, \citenamefont {Cr\'epeau}, \citenamefont
		{Jozsa}, \citenamefont {Peres},\ and\ \citenamefont {Wootters}}]{bennett93}%
	\BibitemOpen
	\bibfield  {author} {\bibinfo {author} {\bibfnamefont {C.~H.}\ \bibnamefont
			{Bennett}}, \bibinfo {author} {\bibfnamefont {G.}~\bibnamefont {Brassard}},
		\bibinfo {author} {\bibfnamefont {C.}~\bibnamefont {Cr\'epeau}}, \bibinfo
		{author} {\bibfnamefont {R.}~\bibnamefont {Jozsa}}, \bibinfo {author}
		{\bibfnamefont {A.}~\bibnamefont {Peres}},\ and\ \bibinfo {author}
		{\bibfnamefont {W.~K.}\ \bibnamefont {Wootters}},\ }\bibfield  {title}
	{\bibinfo {title} {Teleporting an unknown quantum state via dual classical
			and {Einstein-Podolsky-Rosen} channels},\ }\href
	{https://doi.org/10.1103/PhysRevLett.70.1895} {\bibfield  {journal} {\bibinfo
			{journal} {Phys. Rev. Lett.}\ }\textbf {\bibinfo {volume} {70}},\ \bibinfo
		{pages} {1895} (\bibinfo {year} {1993})}\BibitemShut {NoStop}%
	\bibitem [{\citenamefont {Bouwmeester}\ \emph {et~al.}(1997)\citenamefont
		{Bouwmeester}, \citenamefont {Pan}, \citenamefont {Mattle}, \citenamefont
		{Eibl}, \citenamefont {Weinfurter},\ and\ \citenamefont {Zeilinger}}]{bou97}%
	\BibitemOpen
	\bibfield  {author} {\bibinfo {author} {\bibfnamefont {D.}~\bibnamefont
			{Bouwmeester}}, \bibinfo {author} {\bibfnamefont {J.-W.}\ \bibnamefont
			{Pan}}, \bibinfo {author} {\bibfnamefont {K.}~\bibnamefont {Mattle}},
		\bibinfo {author} {\bibfnamefont {M.}~\bibnamefont {Eibl}}, \bibinfo {author}
		{\bibfnamefont {H.}~\bibnamefont {Weinfurter}},\ and\ \bibinfo {author}
		{\bibfnamefont {A.}~\bibnamefont {Zeilinger}},\ }\bibfield  {title} {\bibinfo
		{title} {Experimental quantum teleportation},\ }\href@noop {} {\bibfield
		{journal} {\bibinfo  {journal} {Nature}\ }\textbf {\bibinfo {volume} {390}},\
		\bibinfo {pages} {575} (\bibinfo {year} {1997})}\BibitemShut {NoStop}%
	\bibitem [{\citenamefont {Vaidman}(1994)}]{vaid94}%
	\BibitemOpen
	\bibfield  {author} {\bibinfo {author} {\bibfnamefont {L.}~\bibnamefont
			{Vaidman}},\ }\bibfield  {title} {\bibinfo {title} {Teleportation of quantum
			states},\ }\href {https://doi.org/10.1103/PhysRevA.49.1473} {\bibfield
		{journal} {\bibinfo  {journal} {Phys. Rev. A}\ }\textbf {\bibinfo {volume}
			{49}},\ \bibinfo {pages} {1473} (\bibinfo {year} {1994})}\BibitemShut
	{NoStop}%
	\bibitem [{\citenamefont {Braunstein}\ and\ \citenamefont
		{Kimble}(1998)}]{bra98}%
	\BibitemOpen
	\bibfield  {author} {\bibinfo {author} {\bibfnamefont {S.~L.}\ \bibnamefont
			{Braunstein}}\ and\ \bibinfo {author} {\bibfnamefont {H.~J.}\ \bibnamefont
			{Kimble}},\ }\bibfield  {title} {\bibinfo {title} {Teleportation of
			continuous quantum variables},\ }\href
	{https://doi.org/10.1103/PhysRevLett.80.869} {\bibfield  {journal} {\bibinfo
			{journal} {Phys. Rev. Lett.}\ }\textbf {\bibinfo {volume} {80}},\ \bibinfo
		{pages} {869} (\bibinfo {year} {1998})}\BibitemShut {NoStop}%
	\bibitem [{\citenamefont {Furusawa}\ \emph
		{et~al.}(1998{\natexlab{a}})\citenamefont {Furusawa}, \citenamefont
		{Sørensen}, \citenamefont {Braunstein}, \citenamefont {Fuchs}, \citenamefont
		{Kimble},\ and\ \citenamefont {Polzik}}]{furu98}%
	\BibitemOpen
	\bibfield  {author} {\bibinfo {author} {\bibfnamefont {A.}~\bibnamefont
			{Furusawa}}, \bibinfo {author} {\bibfnamefont {J.~L.}\ \bibnamefont
			{Sørensen}}, \bibinfo {author} {\bibfnamefont {S.~L.}\ \bibnamefont
			{Braunstein}}, \bibinfo {author} {\bibfnamefont {C.~A.}\ \bibnamefont
			{Fuchs}}, \bibinfo {author} {\bibfnamefont {H.~J.}\ \bibnamefont {Kimble}},\
		and\ \bibinfo {author} {\bibfnamefont {E.~S.}\ \bibnamefont {Polzik}},\
	}\bibfield  {title} {\bibinfo {title} {Unconditional quantum teleportation},\
	}\href@noop {} {\bibfield  {journal} {\bibinfo  {journal} {Science}\ }\textbf
		{\bibinfo {volume} {282}},\ \bibinfo {pages} {706} (\bibinfo {year}
		{1998}{\natexlab{a}})}\BibitemShut {NoStop}%
	\bibitem [{\citenamefont {Boto}\ \emph {et~al.}(2000)\citenamefont {Boto},
		\citenamefont {Kok}, \citenamefont {Abrams}, \citenamefont {Braunstein},
		\citenamefont {Williams},\ and\ \citenamefont {Dowling}}]{dowl00}%
	\BibitemOpen
	\bibfield  {author} {\bibinfo {author} {\bibfnamefont {A.~N.}\ \bibnamefont
			{Boto}}, \bibinfo {author} {\bibfnamefont {P.}~\bibnamefont {Kok}}, \bibinfo
		{author} {\bibfnamefont {D.~S.}\ \bibnamefont {Abrams}}, \bibinfo {author}
		{\bibfnamefont {S.~L.}\ \bibnamefont {Braunstein}}, \bibinfo {author}
		{\bibfnamefont {C.~P.}\ \bibnamefont {Williams}},\ and\ \bibinfo {author}
		{\bibfnamefont {J.~P.}\ \bibnamefont {Dowling}},\ }\bibfield  {title}
	{\bibinfo {title} {Quantum interferometric optical lithography: Exploiting
			entanglement to beat the diffraction limit},\ }\href
	{https://doi.org/10.1103/PhysRevLett.85.2733} {\bibfield  {journal} {\bibinfo
			{journal} {Phys. Rev. Lett.}\ }\textbf {\bibinfo {volume} {85}},\ \bibinfo
		{pages} {2733} (\bibinfo {year} {2000})}\BibitemShut {NoStop}%
	\bibitem [{\citenamefont {Giovannetti}\ \emph {et~al.}(2006)\citenamefont
		{Giovannetti}, \citenamefont {Lloyd},\ and\ \citenamefont {Maccone}}]{gio06}%
	\BibitemOpen
	\bibfield  {author} {\bibinfo {author} {\bibfnamefont {V.}~\bibnamefont
			{Giovannetti}}, \bibinfo {author} {\bibfnamefont {S.}~\bibnamefont {Lloyd}},\
		and\ \bibinfo {author} {\bibfnamefont {L.}~\bibnamefont {Maccone}},\
	}\bibfield  {title} {\bibinfo {title} {Quantum metrology},\ }\href
	{https://doi.org/10.1103/PhysRevLett.96.010401} {\bibfield  {journal}
		{\bibinfo  {journal} {Phys. Rev. Lett.}\ }\textbf {\bibinfo {volume} {96}},\
		\bibinfo {pages} {010401} (\bibinfo {year} {2006})}\BibitemShut {NoStop}%
	\bibitem [{\citenamefont {Ma}\ \emph {et~al.}(2017)\citenamefont {Ma},
		\citenamefont {Miao}, \citenamefont {Pang}, \citenamefont {Evans},
		\citenamefont {Zhao}, \citenamefont {Harms}, \citenamefont {Schnabel},\ and\
		\citenamefont {Chen}}]{ligoepr}%
	\BibitemOpen
	\bibfield  {author} {\bibinfo {author} {\bibfnamefont {Y.}~\bibnamefont
			{Ma}}, \bibinfo {author} {\bibfnamefont {H.}~\bibnamefont {Miao}}, \bibinfo
		{author} {\bibfnamefont {B.~H.}\ \bibnamefont {Pang}}, \bibinfo {author}
		{\bibfnamefont {M.}~\bibnamefont {Evans}}, \bibinfo {author} {\bibfnamefont
			{C.}~\bibnamefont {Zhao}}, \bibinfo {author} {\bibfnamefont {J.}~\bibnamefont
			{Harms}}, \bibinfo {author} {\bibfnamefont {R.}~\bibnamefont {Schnabel}},\
		and\ \bibinfo {author} {\bibfnamefont {Y.}~\bibnamefont {Chen}},\ }\bibfield
	{title} {\bibinfo {title} {Proposal for gravitational-wave detection beyond
			the standard quantum limit through {EPR} entanglement},\ }\href@noop {}
	{\bibfield  {journal} {\bibinfo  {journal} {Nature Physics}\ }\textbf
		{\bibinfo {volume} {13}},\ \bibinfo {pages} {776} (\bibinfo {year}
		{2017})}\BibitemShut {NoStop}%
	\bibitem [{\citenamefont {Shaked}\ \emph {et~al.}(2018)\citenamefont {Shaked},
		\citenamefont {Michael}, \citenamefont {Vered}, \citenamefont {Bello},
		\citenamefont {Rosenbluh},\ and\ \citenamefont {Pe'er}}]{Shaked-NC}%
	\BibitemOpen
	\bibfield  {author} {\bibinfo {author} {\bibfnamefont {Y.}~\bibnamefont
			{Shaked}}, \bibinfo {author} {\bibfnamefont {Y.}~\bibnamefont {Michael}},
		\bibinfo {author} {\bibfnamefont {R.~Z.}\ \bibnamefont {Vered}}, \bibinfo
		{author} {\bibfnamefont {L.}~\bibnamefont {Bello}}, \bibinfo {author}
		{\bibfnamefont {M.}~\bibnamefont {Rosenbluh}},\ and\ \bibinfo {author}
		{\bibfnamefont {A.}~\bibnamefont {Pe'er}},\ }\bibfield  {title} {\bibinfo
		{title} {Lifting the bandwidth limit of optical homodyne measurement with
			broadband parametric amplification},\ }\href@noop {} {\bibfield  {journal}
		{\bibinfo  {journal} {Nat Commun}\ }\textbf {\bibinfo {volume} {9}},\
		\bibinfo {pages} {609} (\bibinfo {year} {2018})}\BibitemShut {NoStop}%
	\bibitem [{\citenamefont {Flurin}\ \emph {et~al.}(2012)\citenamefont {Flurin},
		\citenamefont {Roch}, \citenamefont {Mallet}, \citenamefont {Devoret},\ and\
		\citenamefont {Huard}}]{Flurin-PRL}%
	\BibitemOpen
	\bibfield  {author} {\bibinfo {author} {\bibfnamefont {E.}~\bibnamefont
			{Flurin}}, \bibinfo {author} {\bibfnamefont {N.}~\bibnamefont {Roch}},
		\bibinfo {author} {\bibfnamefont {F.}~\bibnamefont {Mallet}}, \bibinfo
		{author} {\bibfnamefont {M.~H.}\ \bibnamefont {Devoret}},\ and\ \bibinfo
		{author} {\bibfnamefont {B.}~\bibnamefont {Huard}},\ }\bibfield  {title}
	{\bibinfo {title} {Generating entangled microwave radiation over two
			transmission lines},\ }\href@noop {} {\bibfield  {journal} {\bibinfo
			{journal} {Phys Rev Lett}\ }\textbf {\bibinfo {volume} {109}},\ \bibinfo
		{pages} {183901} (\bibinfo {year} {2012})}\BibitemShut {NoStop}%
	\bibitem [{\citenamefont {Einstein}\ \emph {et~al.}(1935)\citenamefont
		{Einstein}, \citenamefont {Podolsky},\ and\ \citenamefont {Rosen}}]{epr}%
	\BibitemOpen
	\bibfield  {author} {\bibinfo {author} {\bibfnamefont {A.}~\bibnamefont
			{Einstein}}, \bibinfo {author} {\bibfnamefont {B.}~\bibnamefont {Podolsky}},\
		and\ \bibinfo {author} {\bibfnamefont {N.}~\bibnamefont {Rosen}},\ }\bibfield
	{title} {\bibinfo {title} {Can quantum-mechanical description of physical
			reality be considered complete?},\ }\href
	{https://doi.org/10.1103/PhysRev.47.777} {\bibfield  {journal} {\bibinfo
			{journal} {Phys. Rev.}\ }\textbf {\bibinfo {volume} {47}},\ \bibinfo {pages}
		{777} (\bibinfo {year} {1935})}\BibitemShut {NoStop}%
	\bibitem [{\citenamefont {Reid}(1989)}]{reid89}%
	\BibitemOpen
	\bibfield  {author} {\bibinfo {author} {\bibfnamefont {M.~D.}\ \bibnamefont
			{Reid}},\ }\bibfield  {title} {\bibinfo {title} {Demonstration of the
			{Einstein-Podolsky-Rosen} paradox using nondegenerate parametric
			amplification},\ }\href {https://doi.org/10.1103/PhysRevA.40.913} {\bibfield
		{journal} {\bibinfo  {journal} {Phys. Rev. A}\ }\textbf {\bibinfo {volume}
			{40}},\ \bibinfo {pages} {913} (\bibinfo {year} {1989})}\BibitemShut
	{NoStop}%
	\bibitem [{\citenamefont {Ou}\ \emph {et~al.}(1992)\citenamefont {Ou},
		\citenamefont {Pereira}, \citenamefont {Kimble},\ and\ \citenamefont
		{Peng}}]{ou92}%
	\BibitemOpen
	\bibfield  {author} {\bibinfo {author} {\bibfnamefont {Z.~Y.}\ \bibnamefont
			{Ou}}, \bibinfo {author} {\bibfnamefont {S.~F.}\ \bibnamefont {Pereira}},
		\bibinfo {author} {\bibfnamefont {H.~J.}\ \bibnamefont {Kimble}},\ and\
		\bibinfo {author} {\bibfnamefont {K.~C.}\ \bibnamefont {Peng}},\ }\bibfield
	{title} {\bibinfo {title} {Realization of the einstein-podolsky-rosen paradox
			for continuous variables},\ }\href@noop {} {\bibfield  {journal} {\bibinfo
			{journal} {Phys. Rev. Lett.}\ }\textbf {\bibinfo {volume} {68}},\ \bibinfo
		{pages} {3663} (\bibinfo {year} {1992})}\BibitemShut {NoStop}%
	\bibitem [{\citenamefont {Duan}\ \emph {et~al.}(2000)\citenamefont {Duan},
		\citenamefont {Giedke}, \citenamefont {Cirac},\ and\ \citenamefont
		{Zoller}}]{duan00}%
	\BibitemOpen
	\bibfield  {author} {\bibinfo {author} {\bibfnamefont {L.-M.}\ \bibnamefont
			{Duan}}, \bibinfo {author} {\bibfnamefont {G.}~\bibnamefont {Giedke}},
		\bibinfo {author} {\bibfnamefont {J.~I.}\ \bibnamefont {Cirac}},\ and\
		\bibinfo {author} {\bibfnamefont {P.}~\bibnamefont {Zoller}},\ }\bibfield
	{title} {\bibinfo {title} {Inseparability criterion for continuous variable
			systems},\ }\href {https://doi.org/10.1103/PhysRevLett.84.2722} {\bibfield
		{journal} {\bibinfo  {journal} {Phys. Rev. Lett.}\ }\textbf {\bibinfo
			{volume} {84}},\ \bibinfo {pages} {2722} (\bibinfo {year}
		{2000})}\BibitemShut {NoStop}%
	\bibitem [{\citenamefont {Slusher}\ \emph {et~al.}(1987)\citenamefont
		{Slusher}, \citenamefont {Grangier}, \citenamefont {LaPorta}, \citenamefont
		{Yurke},\ and\ \citenamefont {Potasek}}]{Slusher87}%
	\BibitemOpen
	\bibfield  {author} {\bibinfo {author} {\bibfnamefont {R.~E.}\ \bibnamefont
			{Slusher}}, \bibinfo {author} {\bibfnamefont {P.}~\bibnamefont {Grangier}},
		\bibinfo {author} {\bibfnamefont {A.}~\bibnamefont {LaPorta}}, \bibinfo
		{author} {\bibfnamefont {B.}~\bibnamefont {Yurke}},\ and\ \bibinfo {author}
		{\bibfnamefont {M.~J.}\ \bibnamefont {Potasek}},\ }\bibfield  {title}
	{\bibinfo {title} {Pulsed squeezed light},\ }\href
	{https://doi.org/10.1103/PhysRevLett.59.2566} {\bibfield  {journal} {\bibinfo
			{journal} {Physical Review Letters}\ }\textbf {\bibinfo {volume} {59}},\
		\bibinfo {pages} {2566} (\bibinfo {year} {1987})}\BibitemShut {NoStop}%
	\bibitem [{\citenamefont {Ayt\"ur}\ and\ \citenamefont
		{Kumar}(1990)}]{Kumar90}%
	\BibitemOpen
	\bibfield  {author} {\bibinfo {author} {\bibfnamefont {O.}~\bibnamefont
			{Ayt\"ur}}\ and\ \bibinfo {author} {\bibfnamefont {P.}~\bibnamefont
			{Kumar}},\ }\bibfield  {title} {\bibinfo {title} {Pulsed twin beams of
			light},\ }\href {https://doi.org/10.1103/PhysRevLett.65.1551} {\bibfield
		{journal} {\bibinfo  {journal} {Physical Review Letters}\ }\textbf {\bibinfo
			{volume} {65}},\ \bibinfo {pages} {1551} (\bibinfo {year}
		{1990})}\BibitemShut {NoStop}%
	\bibitem [{\citenamefont {Wenger}\ \emph {et~al.}(2004)\citenamefont {Wenger},
		\citenamefont {Tualle-Brouri},\ and\ \citenamefont {Grangier}}]{Wenger04}%
	\BibitemOpen
	\bibfield  {author} {\bibinfo {author} {\bibfnamefont {J.}~\bibnamefont
			{Wenger}}, \bibinfo {author} {\bibfnamefont {R.}~\bibnamefont
			{Tualle-Brouri}},\ and\ \bibinfo {author} {\bibfnamefont {P.}~\bibnamefont
			{Grangier}},\ }\bibfield  {title} {\bibinfo {title} {Pulsed homodyne
			measurements of femtosecond squeezed pulses generated by single-pass
			parametric deamplification},\ }\href@noop {} {\bibfield  {journal} {\bibinfo
			{journal} {Optics Letters}\ }\textbf {\bibinfo {volume} {29}},\ \bibinfo
		{pages} {1267} (\bibinfo {year} {2004})}\BibitemShut {NoStop}%
	\bibitem [{\citenamefont {Eto}\ \emph {et~al.}(2008)\citenamefont {Eto},
		\citenamefont {Tajima}, \citenamefont {Zhang},\ and\ \citenamefont
		{Hirano}}]{Eto08}%
	\BibitemOpen
	\bibfield  {author} {\bibinfo {author} {\bibfnamefont {Y.}~\bibnamefont
			{Eto}}, \bibinfo {author} {\bibfnamefont {T.}~\bibnamefont {Tajima}},
		\bibinfo {author} {\bibfnamefont {Y.}~\bibnamefont {Zhang}},\ and\ \bibinfo
		{author} {\bibfnamefont {T.}~\bibnamefont {Hirano}},\ }\bibfield  {title}
	{\bibinfo {title} {Observation of quadrature squeezing in a $\chi^{(2)}$
			nonlinear waveguide using a temporally shaped local oscillator pulse},\
	}\href@noop {} {\bibfield  {journal} {\bibinfo  {journal} {Optics Express}\
		}\textbf {\bibinfo {volume} {16}},\ \bibinfo {pages} {10650} (\bibinfo {year}
		{2008})}\BibitemShut {NoStop}%
	\bibitem [{\citenamefont {Guo}\ \emph {et~al.}(2016)\citenamefont {Guo},
		\citenamefont {Liu}, \citenamefont {Liu}, \citenamefont {Li},\ and\
		\citenamefont {Ou}}]{Guo16}%
	\BibitemOpen
	\bibfield  {author} {\bibinfo {author} {\bibfnamefont {X.}~\bibnamefont
			{Guo}}, \bibinfo {author} {\bibfnamefont {N.}~\bibnamefont {Liu}}, \bibinfo
		{author} {\bibfnamefont {Y.}~\bibnamefont {Liu}}, \bibinfo {author}
		{\bibfnamefont {X.}~\bibnamefont {Li}},\ and\ \bibinfo {author}
		{\bibfnamefont {Z.~Y.}\ \bibnamefont {Ou}},\ }\bibfield  {title} {\bibinfo
		{title} {Generation of continuous variable quantum entanglement using a fiber
			optical parametric amplifier},\ }\href@noop {} {\bibfield  {journal}
		{\bibinfo  {journal} {Optics letters}\ }\textbf {\bibinfo {volume} {41}},\
		\bibinfo {pages} {653} (\bibinfo {year} {2016})}\BibitemShut {NoStop}%
	\bibitem [{\citenamefont {Wasilewski}\ \emph {et~al.}(2006)\citenamefont
		{Wasilewski}, \citenamefont {Lvovsky}, \citenamefont {Banaszek},\ and\
		\citenamefont {Radzewicz}}]{Wasilewski06}%
	\BibitemOpen
	\bibfield  {author} {\bibinfo {author} {\bibfnamefont {W.}~\bibnamefont
			{Wasilewski}}, \bibinfo {author} {\bibfnamefont {A.~I.}\ \bibnamefont
			{Lvovsky}}, \bibinfo {author} {\bibfnamefont {K.}~\bibnamefont {Banaszek}},\
		and\ \bibinfo {author} {\bibfnamefont {C.}~\bibnamefont {Radzewicz}},\
	}\bibfield  {title} {\bibinfo {title} {Pulsed squeezed light: Simultaneous
			squeezing of multiple modes},\ }\href
	{https://doi.org/10.1103/PhysRevA.73.063819} {\bibfield  {journal} {\bibinfo
			{journal} {Physical Review A}\ }\textbf {\bibinfo {volume} {73}},\ \bibinfo
		{pages} {063819} (\bibinfo {year} {2006})}\BibitemShut {NoStop}%
	\bibitem [{\citenamefont {Guo}\ \emph {et~al.}(2015)\citenamefont {Guo},
		\citenamefont {Liu}, \citenamefont {Li},\ and\ \citenamefont {Ou}}]{guo15}%
	\BibitemOpen
	\bibfield  {author} {\bibinfo {author} {\bibfnamefont {X.}~\bibnamefont
			{Guo}}, \bibinfo {author} {\bibfnamefont {N.}~\bibnamefont {Liu}}, \bibinfo
		{author} {\bibfnamefont {X.}~\bibnamefont {Li}},\ and\ \bibinfo {author}
		{\bibfnamefont {Z.~Y.}\ \bibnamefont {Ou}},\ }\bibfield  {title} {\bibinfo
		{title} {Complete temporal mode analysis in pulse-pumped fiber-optical
			parametric amplifier for continuous variable entanglement generation},\
	}\href@noop {} {\bibfield  {journal} {\bibinfo  {journal} {Optics Express}\
		}\textbf {\bibinfo {volume} {23}},\ \bibinfo {pages} {29369} (\bibinfo {year}
		{2015})}\BibitemShut {NoStop}%
	\bibitem [{\citenamefont {Xin}\ \emph {et~al.}(2017)\citenamefont {Xin},
		\citenamefont {Qi},\ and\ \citenamefont {Jing}}]{XJ17}%
	\BibitemOpen
	\bibfield  {author} {\bibinfo {author} {\bibfnamefont {J.}~\bibnamefont
			{Xin}}, \bibinfo {author} {\bibfnamefont {J.}~\bibnamefont {Qi}},\ and\
		\bibinfo {author} {\bibfnamefont {J.}~\bibnamefont {Jing}},\ }\bibfield
	{title} {\bibinfo {title} {Enhancement of entanglement using cascaded
			four-wave mixing processes},\ }\href@noop {} {\bibfield  {journal} {\bibinfo
			{journal} {Optics Letters}\ }\textbf {\bibinfo {volume} {42}},\ \bibinfo
		{pages} {366} (\bibinfo {year} {2017})}\BibitemShut {NoStop}%
	\bibitem [{\citenamefont {Furusawa}\ \emph
		{et~al.}(1998{\natexlab{b}})\citenamefont {Furusawa}, \citenamefont
		{Sørensen}, \citenamefont {Braunstein}, \citenamefont {Fuchs}, \citenamefont
		{Kimble},\ and\ \citenamefont {Polzik}}]{Furusawa98}%
	\BibitemOpen
	\bibfield  {author} {\bibinfo {author} {\bibfnamefont {A.}~\bibnamefont
			{Furusawa}}, \bibinfo {author} {\bibfnamefont {J.~L.}\ \bibnamefont
			{Sørensen}}, \bibinfo {author} {\bibfnamefont {S.~L.}\ \bibnamefont
			{Braunstein}}, \bibinfo {author} {\bibfnamefont {C.~A.}\ \bibnamefont
			{Fuchs}}, \bibinfo {author} {\bibfnamefont {H.~J.}\ \bibnamefont {Kimble}},\
		and\ \bibinfo {author} {\bibfnamefont {E.~S.}\ \bibnamefont {Polzik}},\
	}\bibfield  {title} {\bibinfo {title} {Unconditional quantum teleportation},\
	}\href@noop {} {\bibfield  {journal} {\bibinfo  {journal} {Science}\ }\textbf
		{\bibinfo {volume} {282}},\ \bibinfo {pages} {706} (\bibinfo {year}
		{1998}{\natexlab{b}})}\BibitemShut {NoStop}%
	\bibitem [{\citenamefont {Li}\ \emph {et~al.}(2018)\citenamefont {Li},
		\citenamefont {Liu}, \citenamefont {Cui}, \citenamefont {Huo}, \citenamefont
		{Assad}, \citenamefont {Li},\ and\ \citenamefont {Ou}}]{JM-PRA18}%
	\BibitemOpen
	\bibfield  {author} {\bibinfo {author} {\bibfnamefont {J.}~\bibnamefont
			{Li}}, \bibinfo {author} {\bibfnamefont {Y.}~\bibnamefont {Liu}}, \bibinfo
		{author} {\bibfnamefont {L.}~\bibnamefont {Cui}}, \bibinfo {author}
		{\bibfnamefont {N.}~\bibnamefont {Huo}}, \bibinfo {author} {\bibfnamefont
			{S.~M.}\ \bibnamefont {Assad}}, \bibinfo {author} {\bibfnamefont
			{X.}~\bibnamefont {Li}},\ and\ \bibinfo {author} {\bibfnamefont {Z.~Y.}\
			\bibnamefont {Ou}},\ }\bibfield  {title} {\bibinfo {title} {Joint measurement
			of multiple noncommuting parameters},\ }\href@noop {} {\bibfield  {journal}
		{\bibinfo  {journal} {Physical Review A}\ }\textbf {\bibinfo {volume} {97}},\
		\bibinfo {pages} {052127} (\bibinfo {year} {2018})}\BibitemShut {NoStop}%
	\bibitem [{\citenamefont {Ou}(2012)}]{Ou12}%
	\BibitemOpen
	\bibfield  {author} {\bibinfo {author} {\bibfnamefont {Z.~Y.}\ \bibnamefont
			{Ou}},\ }\bibfield  {title} {\bibinfo {title} {Enhancement of the
			phase-measurement sensitivity beyond the standard quantum limit by a
			nonlinear interferometer},\ }\href@noop {} {\bibfield  {journal} {\bibinfo
			{journal} {Physical Review A}\ }\textbf {\bibinfo {volume} {85}} (\bibinfo
		{year} {2012})}\BibitemShut {NoStop}%
	\bibitem [{\citenamefont {Christ}\ \emph {et~al.}(2011)\citenamefont {Christ},
		\citenamefont {Laiho}, \citenamefont {Eckstein}, \citenamefont {Cassemiro},\
		and\ \citenamefont {Silberhorn}}]{sil}%
	\BibitemOpen
	\bibfield  {author} {\bibinfo {author} {\bibfnamefont {A.}~\bibnamefont
			{Christ}}, \bibinfo {author} {\bibfnamefont {K.}~\bibnamefont {Laiho}},
		\bibinfo {author} {\bibfnamefont {A.}~\bibnamefont {Eckstein}}, \bibinfo
		{author} {\bibfnamefont {K.~N.}\ \bibnamefont {Cassemiro}},\ and\ \bibinfo
		{author} {\bibfnamefont {C.}~\bibnamefont {Silberhorn}},\ }\bibfield  {title}
	{\bibinfo {title} {Probing multimode squeezing with correlation functions},\
	}\href@noop {} {\bibfield  {journal} {\bibinfo  {journal} {New Journal of
				Physics}\ }\textbf {\bibinfo {volume} {13}},\ \bibinfo {pages} {033027}
		(\bibinfo {year} {2011})}\BibitemShut {NoStop}%
	\bibitem [{\citenamefont {Liu}\ \emph {et~al.}(2016)\citenamefont {Liu},
		\citenamefont {Liu}, \citenamefont {Guo}, \citenamefont {Yang}, \citenamefont
		{Li*},\ and\ \citenamefont {Ou}}]{LNN16}%
	\BibitemOpen
	\bibfield  {author} {\bibinfo {author} {\bibfnamefont {N.}~\bibnamefont
			{Liu}}, \bibinfo {author} {\bibfnamefont {Y.}~\bibnamefont {Liu}}, \bibinfo
		{author} {\bibfnamefont {X.}~\bibnamefont {Guo}}, \bibinfo {author}
		{\bibfnamefont {L.}~\bibnamefont {Yang}}, \bibinfo {author} {\bibfnamefont
			{X.}~\bibnamefont {Li*}},\ and\ \bibinfo {author} {\bibfnamefont {Z.~Y.}\
			\bibnamefont {Ou}},\ }\bibfield  {title} {\bibinfo {title} {Approaching
			single temporal mode operation in twin beams generated by pulse pumped high
			gain spontaneous four wave mixing},\ }\href@noop {} {\bibfield  {journal}
		{\bibinfo  {journal} {Optics Express}\ }\textbf {\bibinfo {volume} {24}},\
		\bibinfo {pages} {1096} (\bibinfo {year} {2016})}\BibitemShut {NoStop}%
	\bibitem [{\citenamefont {Li}\ \emph {et~al.}(2019)\citenamefont {Li},
		\citenamefont {Liu}, \citenamefont {Huo}, \citenamefont {Cui}, \citenamefont
		{Feng}, \citenamefont {Ou},\ and\ \citenamefont {Li}}]{LJM19}%
	\BibitemOpen
	\bibfield  {author} {\bibinfo {author} {\bibfnamefont {J.}~\bibnamefont
			{Li}}, \bibinfo {author} {\bibfnamefont {Y.}~\bibnamefont {Liu}}, \bibinfo
		{author} {\bibfnamefont {N.}~\bibnamefont {Huo}}, \bibinfo {author}
		{\bibfnamefont {L.}~\bibnamefont {Cui}}, \bibinfo {author} {\bibfnamefont
			{C.}~\bibnamefont {Feng}}, \bibinfo {author} {\bibfnamefont {Z.~Y.}\
			\bibnamefont {Ou}},\ and\ \bibinfo {author} {\bibfnamefont {X.}~\bibnamefont
			{Li}},\ }\bibfield  {title} {\bibinfo {title} {Pulsed entanglement measured
			by parametric amplifier assisted homodyne detection},\ }\href@noop {}
	{\bibfield  {journal} {\bibinfo  {journal} {Optics Express}\ }\textbf
		{\bibinfo {volume} {27}},\ \bibinfo {pages} {30552} (\bibinfo {year}
		{2019})}\BibitemShut {NoStop}%
	\bibitem [{\citenamefont {Liu}\ \emph {et~al.}(2019)\citenamefont {Liu},
		\citenamefont {Huo}, \citenamefont {Li}, \citenamefont {Cui}, \citenamefont
		{Li},\ and\ \citenamefont {Ou}}]{liu-OE19}%
	\BibitemOpen
	\bibfield  {author} {\bibinfo {author} {\bibfnamefont {Y.}~\bibnamefont
			{Liu}}, \bibinfo {author} {\bibfnamefont {N.}~\bibnamefont {Huo}}, \bibinfo
		{author} {\bibfnamefont {J.}~\bibnamefont {Li}}, \bibinfo {author}
		{\bibfnamefont {L.}~\bibnamefont {Cui}}, \bibinfo {author} {\bibfnamefont
			{X.}~\bibnamefont {Li}},\ and\ \bibinfo {author} {\bibfnamefont {Z.~J.}\
			\bibnamefont {Ou}},\ }\bibfield  {title} {\bibinfo {title} {Optimum quantum
			resource distribution for phase measurement and quantum information tapping
			in a dual-beam su(1,1) interferometer},\ }\href@noop {} {\bibfield  {journal}
		{\bibinfo  {journal} {Optics Express}\ }\textbf {\bibinfo {volume} {27}},\
		\bibinfo {pages} {11292} (\bibinfo {year} {2019})}\BibitemShut {NoStop}%
	\bibitem [{\citenamefont {Liu}\ \emph {et~al.}(2018)\citenamefont {Liu},
		\citenamefont {Li}, \citenamefont {Cui}, \citenamefont {Huo}, \citenamefont
		{Assad}, \citenamefont {Li},\ and\ \citenamefont {Ou}}]{liu-OE18}%
	\BibitemOpen
	\bibfield  {author} {\bibinfo {author} {\bibfnamefont {Y.}~\bibnamefont
			{Liu}}, \bibinfo {author} {\bibfnamefont {J.}~\bibnamefont {Li}}, \bibinfo
		{author} {\bibfnamefont {L.}~\bibnamefont {Cui}}, \bibinfo {author}
		{\bibfnamefont {N.}~\bibnamefont {Huo}}, \bibinfo {author} {\bibfnamefont
			{S.~M.}\ \bibnamefont {Assad}}, \bibinfo {author} {\bibfnamefont
			{X.}~\bibnamefont {Li}},\ and\ \bibinfo {author} {\bibfnamefont {Z.~Y.}\
			\bibnamefont {Ou}},\ }\bibfield  {title} {\bibinfo {title} {Loss-tolerant
			quantum dense metrology with su(1,1) interferometer},\ }\href@noop {}
	{\bibfield  {journal} {\bibinfo  {journal} {Optics Express}\ }\textbf
		{\bibinfo {volume} {26}},\ \bibinfo {pages} {27705} (\bibinfo {year}
		{2018})}\BibitemShut {NoStop}%
	\bibitem [{\citenamefont {Zhang}\ \emph {et~al.}(2000)\citenamefont {Zhang},
		\citenamefont {Wang}, \citenamefont {Li}, \citenamefont {Jing}, \citenamefont
		{Xie},\ and\ \citenamefont {Peng*}}]{ZhY00}%
	\BibitemOpen
	\bibfield  {author} {\bibinfo {author} {\bibfnamefont {Y.}~\bibnamefont
			{Zhang}}, \bibinfo {author} {\bibfnamefont {H.}~\bibnamefont {Wang}},
		\bibinfo {author} {\bibfnamefont {X.}~\bibnamefont {Li}}, \bibinfo {author}
		{\bibfnamefont {J.}~\bibnamefont {Jing}}, \bibinfo {author} {\bibfnamefont
			{C.}~\bibnamefont {Xie}},\ and\ \bibinfo {author} {\bibfnamefont
			{K.}~\bibnamefont {Peng*}},\ }\bibfield  {title} {\bibinfo {title}
		{Experimental generation of bright two-mode quadrature squeezed light from a
			narrow-band nondegenerate optical parametric amplifier},\ }\href@noop {}
	{\bibfield  {journal} {\bibinfo  {journal} {Physical Review A}\ }\textbf
		{\bibinfo {volume} {62}},\ \bibinfo {pages} {023813} (\bibinfo {year}
		{2000})}\BibitemShut {NoStop}%
\end{thebibliography}
%

\end{document}